\documentclass[aps,prl,groupedaddress,twocolumn]{revtex4-1}
\usepackage[dvips]{graphicx}
\usepackage{dcolumn}
\usepackage{bm}
\usepackage{amsmath,amsthm,amssymb}
\usepackage{color}
\usepackage{verbatim}
\usepackage{wasysym}

\usepackage[colorinlistoftodos,textsize=small,textwidth=columnwidth]{todonotes} 

\newcommand*\circled[1]{\tikz[baseline=(char.base)]{
  \node[shape=circle,draw,inner sep=1pt] (char) {#1};}}
  
\newcommand{\ket}[1]{\left| #1 \right>} 

\begin{document}

\title{An experimental approach for investigating many-body phenomena in Rydberg-interacting quantum systems}
\author{C. S. Hofmann}
\email{chofmann@physi.uni-heidelberg.de}
\author{G. G\"unter}
\author{H. Schempp}
\author{N. L. M. M\"uller}
\altaffiliation{Present address: Center for Free-Electron Laser Science, DESY, Notkestrasse 85, 22607 Hamburg, Germany}
\author{A. Faber}
\altaffiliation{Present address: Department of Physics, University of Basel, Klingelbergstrasse 82, 4056 Basel, Switzerland}
\author{H.~Busche}
\altaffiliation{Present address: Joint Quantum Centre (JQC) Durham-Newcastle, Department of Physics, Durham University, Rochester Building, South Road, Durham DH1 3LE, United Kingdom}
\author{M. {Robert-de-Saint-Vincent}}
\author{S. Whitlock}
\email{whitlock@physi.uni-heidelberg.de}
\author{M. Weidem\"uller}
\email{weidemueller@uni-heidelberg.de}
\affiliation{Physikalisches Institut, Universit\"at Heidelberg, Im Neuenheimer Feld 226, 69120 Heidelberg, Germany.}

\date{\today}

\begin{abstract}
Recent developments in the study of ultracold Rydberg gases demand an advanced level of experimental sophistication, in which high atomic and optical densities must be combined with excellent control of external fields and sensitive Rydberg atom detection. We describe a tailored experimental system used to produce and study Rydberg-interacting atoms excited from dense ultracold atomic gases. The experiment has been optimized for fast duty cycles using a high flux cold atom source and a three beam optical dipole trap. The latter enables tuning of the atomic density and temperature over several orders of magnitude, all the way to the Bose-Einstein condensation transition. An electrode structure surrounding the atoms allows for precise control over electric fields and single-particle sensitive field ionization detection of Rydberg atoms. We review two experiments which highlight the influence of strong Rydberg--Rydberg interactions on different many-body systems. First, the Rydberg blockade effect is used to pre-structure an atomic gas prior to its spontaneous evolution into an ultracold plasma. Second, hybrid states of photons and atoms called dark-state polaritons are studied. By looking at the statistical distribution of Rydberg excited atoms we reveal correlations between dark-state polaritons. These experiments will ultimately provide a deeper understanding of many-body phenomena in strongly-interacting regimes, including the study of strongly-coupled plasmas and interfaces between atoms and light at the quantum level. 
\end{abstract}


\maketitle

\section{I. Introduction}

The experimental and theoretical investigation of ultracold atomic gases involving highly excited Rydberg states has recently attracted
a great deal of interest~\cite{Saffman2010,Comparat2010,Pritchard2013}. One reason is that the strong interactions between Rydberg atoms can dramatically influence the laser excitation process (Rydberg blockade effect), giving rise to intriguing new many-body phases and strong spatial correlations between the atoms ~\cite{Robicheaux2005,Weimer2008,Comparat2010,schwarzkopf2011,Schauss2012,Ates2012,Petrosyan2013}. For example, chirping the excitation laser field may enable the formation of new crystalline states of Rydberg atoms~\cite{Pohl2010,Schachenmayer2010,Vanbijnen2011}. Or, by weakly admixing Rydberg character to ground states with off-resonant laser light it might be possible to introduce new types of interactions between dressed atoms~\cite{santos2000,pupillo2010,henkel2010,henkel2012}.
The Rydberg blockade effect can likewise influence the formation of ultracold plasmas, where spatial correlations between Rydberg atoms prior to their ionization may enable the production of ultracold plasmas in the strongly-coupled regime~\cite{SaintVincent2013,Bannasch2013}. Interestingly, the strong Rydberg--Rydberg interactions can also have a back-action on the laser fields coupling to the Rydberg state. Very recently, electromagnetically induced transparency using Rydberg states was shown to create effective interactions between photons~\cite{Dudin2012,Peyronel2012,Maxwell2013,Hofmann2013}, paving the way to the generation of high-repetition-rate sources of nonclassical light as a resource for quantum information processing or for nonlinear optics involving ultracold Rydberg gases~\cite{Sevinccli2011,Pritchard2013}.

To fully explore these new prospects, modern Rydberg atom experiments require a high level of experimental sophistication in both preparation and detection of the excited atoms. High atomic densities and flexible trapping geometries are a key ingredient, providing large optical densities to enhance nonlinear optical effects, and are used to introduce and control collective many-body effects. Additionally, the development of sensitive detection techniques is required to study the physics of strongly-correlated Rydberg interacting systems, providing direct access to their microscopic details.

In this paper we describe in detail our experimental system tailored for the investigation of strongly-interacting Rydberg gases and the detection of atomic correlations.
In addition to good electric field control, our experiment offers faster experimental cycle times, better magnetic field control, and better optical access as compared to the system recently detailed by R. L\"ow \textit{et al.}~\cite{low2012}. This is possible through the use of a high flux 2D-MOT cold atom source and a three-beam optical dipole trap. The latter allows for varying the atomic density over three-orders of magnitude, including all-optical preparation of $^{87}$Rb Bose-Einstein condensates, and for generating quasi-one-dimensional and three dimensional  Rydberg systems. A special electrode structure is used to control electric fields and to field ionize Rydberg atoms for detection, without limiting the optical access. The paper is structured as follows: Section II contains a detailed description of our experimental apparatus, focusing on the special features used to study Rydberg-interacting quantum systems. In section III we review two experiments performed in the strongly-interacting regime using this apparatus: the spontaneous formation of an ultracold plasma from a Rydberg blockaded gas~\cite{SaintVincent2013}, and the emergence of correlations between dark-state polaritons in Rydberg state electromagnetically induced transparency~\cite{Hofmann2013}.

\section{II. Experiment}

\begin{figure}
  \begin{center}
      \includegraphics[width=0.9\columnwidth]{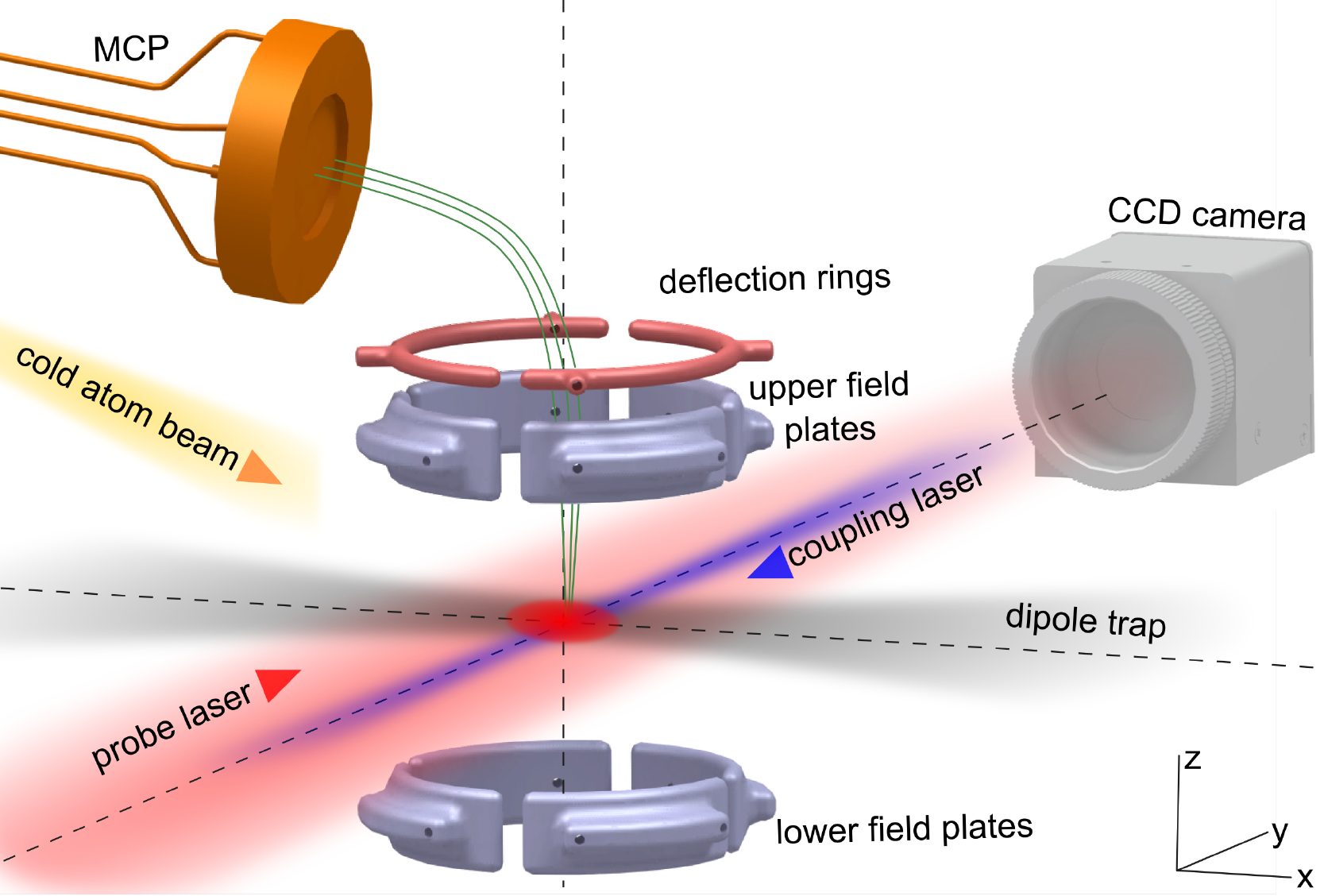}
      \caption{(Color online) \textbf{Experimental system for Rydberg physics in dense ultracold
	       atomic gases.} A high flux cold atom beam loads a magneto-optical trap (not shown).
	       The pre-cooled atoms are subsequently transferred into an optical dipole trap,
	       in which  evaporative cooling is used for precise temperature and density
	       tuning of the gas. After trap release the atoms are excited to high-lying Rydberg
	       states by a $780\,\rm{nm}$ probe laser and a counter-propagating $480\,\rm{nm}$
	       coupling laser. An electrode structure composed of 8 field plates is used for precise
	       electric field control and for field ionization of Rydberg atoms. Two deflection rings
	       are used to  guide the resulting ions (green trajectories) onto a micro-channel
	       plate (MCP) detector. Optical absorption imaging with a CCD camera is done in parallel, 
	       providing complementary information.}
    \label{fig:setup_total}
  \end{center}
\end{figure}

We start by reviewing the technical details of our experimental apparatus. Laser cooling and the used trapping techniques which are common to many ultracold atom experiments are described in detail in reviews by W.~Ketterle \textit{et al.}~\cite{Ketterle1999,Ketterle2008}. Here we focus specifically on the key features of our experimental system (Fig.~\ref{fig:setup_total}) which make it ideally suited for the investigation of Rydberg-interacting quantum systems. The primary goal is to maximize optical access to the atoms, while providing excellent electric field control.  The latter is a main requirement of any Rydberg apparatus, because Rydberg atoms are extremely sensitive to electric fields~\cite{low2012,Sassmannshausen2013}.  

Rydberg atoms  experience large DC-Stark shifts $ \Delta E= -\alpha \mathcal{E}^2/2$, since their atomic polarizability $\alpha$ scales with the seventh power of the principal quantum number $n$. For example, the $\ket{55S_{1/2}}$ state has a polarizability of $\alpha \approx 100\, \rm{MHz}/\rm{(V/cm)}^2$~\cite{OSullivan1985}, while the lifetime limited transition linewidth is only $\approx30\,\rm{kHz}$~\cite{Beterov2009}. Thus, a small electric field of $\mathcal{E} = 1\,\rm{V}/\rm{cm} $ can shift the Rydberg state by several hundred linewidths. On the other hand, moderate electric fields can be used to manipulate Rydberg atom properties, including the use of F\"orster tuning of interactions ~\cite{vogt2006,westermann2006,ryabtsev2010,nipper2012,gurian2012} or spatially resolved excitation using inhomogeneous electric fields~\cite{mulken2007}. Another important consideration is stray fields, caused for example by the presence of a few spurious ions~\cite{Comparat2010}. To minimize this possibility we isolate the cold atom source from the main experimental chamber to eliminate ions emitted from the alkali metal dispensers. For this we use a two-dimensional magneto-optical trap (2D-MOT), which loads a 3D-MOT in the main science chamber~\cite{dieckmann1998,schoser2002,catani2006, chaudhuri2006}. Since the cooling mechanism and hence the transmission through the differential pumping tube are exclusive to the designated neutral atomic species, the source is ion free, while providing high atom fluxes which are comparable to Zeeman slowers. Further advantages of the 2D-MOT are that it does not reduce optical access to the science chamber and that it provides a slow atomic beam with small divergence and high beam luminosity which sets the basis for experiments with high repetition rates.

Once the 3D-MOT is loaded from the cold atom beam, the pre-cooled atoms are directly transfered into an optical dipole trap (ODT). We use all-optical evaporative cooling, in order to precisely control the density and temperature of the atoms and to produce a Bose-Einstein condensate (BEC) if required. Compared to magnetostatic traps~\cite{low2012,dubessy2012}, optical dipole traps are unrestricted to the used spin states and allow for almost instant trap switch-off. Furthermore, they minimally affect the optical access to the atoms and their operation is technically less demanding than hybrid optical and magnetic traps~\cite{altin2010,lin2009}. Finally the realization of fast experimental duty cycles using ODTs is comparable to chip based traps~\cite{Fortagh2007,Reichel2011}.

The good optical access in our experiment is exploited for optical absorption imaging using a charged coupled device (CCD) camera. By imaging the absorption of a weak probe beam transmitted through the atomic cloud in the presence of a second strong coupling laser to the Rydberg states it is possible to observe electromagnetically induced transparency (EIT) in a spatially resolved manner. This serves to map the properties of the Rydberg states onto the probe light field, providing a local probe of the system and the possibility to analyze hundreds of EIT-systems in parallel (mapped onto hundreds of pixels), each with different parameters, within a single experimental run. By combining EIT-based imaging with field ionization detection, one can access both coherences and populations of the interacting many-body system simultaneously~\cite{tobepublished}.

\subsection{A. High flux cold atom source}

\begin{figure}
  \begin{center}
      \includegraphics[width=0.9\columnwidth]{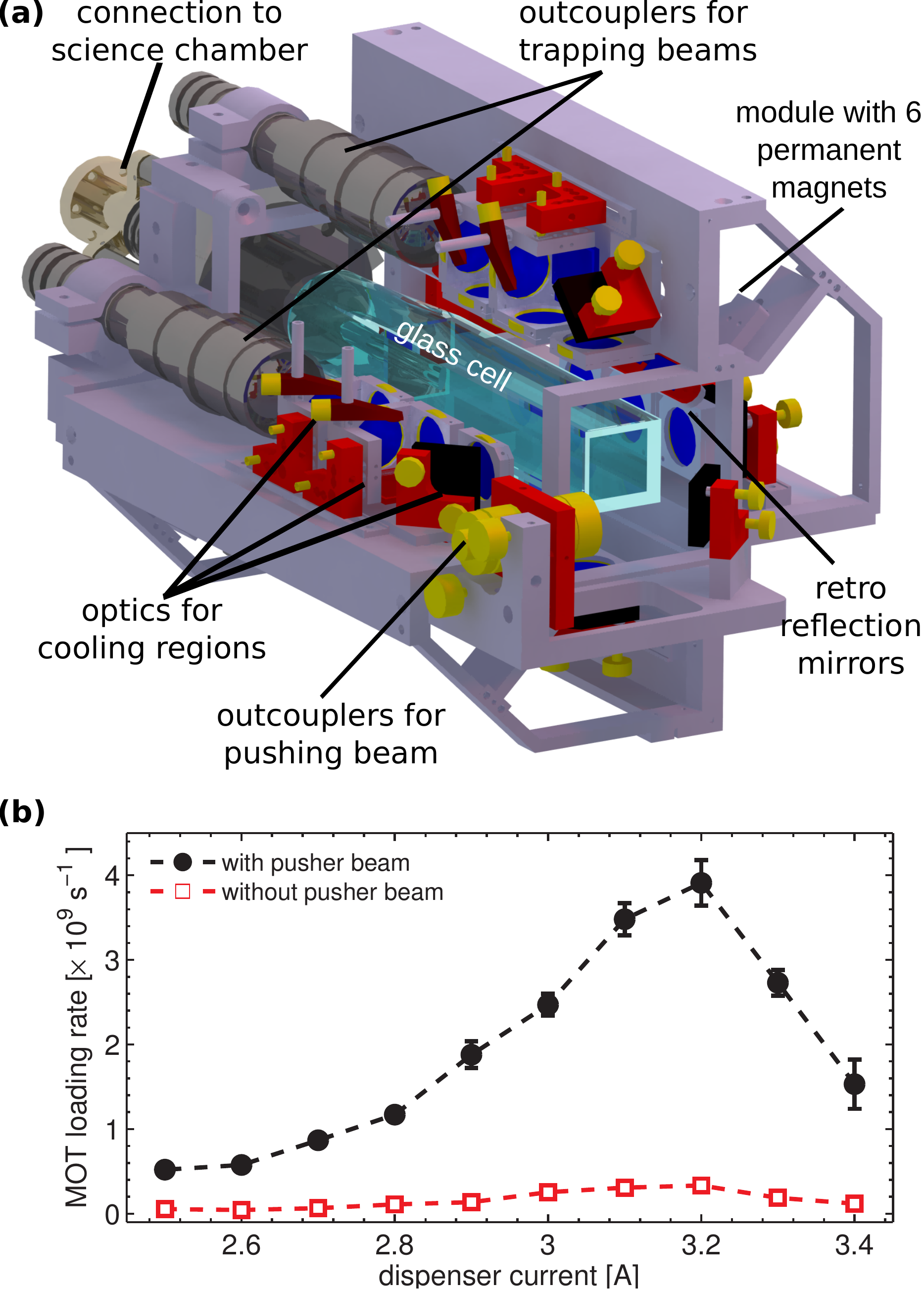}
      \caption{(Color online) \textbf{Technical overview and performance of the 2D-MOT.} (a) The UHV glass cell is surrounded by an attachable and externally adjustable, fiber based 2D-MOT cage system. It provides polarization and retro-reflection optics as well as permanent magnets for three cooling regions. (b) Dependence of the loading rate on the $^{87}\rm{Rb}$ dispenser current. The maximum loading rate of $(3.9\pm0.5)\times 10^9\rm{s}^{-1}$ is reached for a current of $3.2\,\rm{A}$. We observe a loading rate increase of a factor $12\pm2$ when operating the 2D-MOT with pushing beam. The dashed lines serve as a guide to the eye.}
    \label{fig:2dmotmodule}
  \end{center}
\end{figure}

Our 2D-MOT is housed in a $140 \times 35 \times 35\,\rm{mm^3}$ glass cell and is pressure separated from the science chamber by a differential pumping tube. To maintain the
good optical access to the science chamber, we attach the 2D-MOT under an angle of $20^{\circ}$ with respect to the horizontal plane of the experiment. To minimize its space consumption we designed a compact ($200 \times 200 \times 350\,\rm{mm^3}$) fiber based cage system which provides the $780\,\rm{nm}$ cooling light and the magnetic quadrupole fields required for two-dimensional cooling and trapping (see Fig.~\ref{fig:2dmotmodule}(a)).
The cage system can be divided in three planes that intersect in the symmetry axis of the glass cell. Two perpendicular planes hold one inch polarization optics for three equally spaced 2D cooling regions, that are generated by retro-reflected laser beams with $20\,\rm{mm}$ diameter each. Two units with six permanent neodymium bar magnets~\cite{tiecke2009b} are oriented in the third bisecting plane and realize two-dimensional magnetic trapping fields with a field gradient of $16\,\rm{G/cm}$. A $7\,\rm{mm}$ diameter pushing beam propagates along the symmetry axis of the glass cell and guides the pre-cooled atoms through the $800\,\rm{\mu m}$ diameter exit hole of the differential pumping stage into the science chamber. To minimize vacuum contamination in the glass cell, we use $98 \%$ enriched $^{87}\rm{Rb}$ dispensers as our atom source.

For routine operation of the 2D-MOT we use the following parameters which were found to maximize the 3D-MOT loading rate.  The loading rate of the 3D-MOT was measured using absorption images taken at different times during 2D-MOT operation. We typically use $80\,\rm{mW}$ total cooling laser power and $3\,\rm{mW}$ total repumping laser power. The circularly polarized cooling laser is $9\,\rm{MHz}$ red detuned from the $|5S_{1/2}, F=2\rangle \rightarrow |5P_{3/2}, F=3\rangle$ cooling transition and the co-propagating repumping laser is resonant with the $|5S_{1/2}, F=1\rangle \rightarrow |5P_{3/2}, F=2\rangle$ transition. The atom flux is further increased by separately adjusting a pushing beam to direct atoms through the differential pumping tube. The power of the pushing beam is $40\,\rm{mW}$ and the red detuning with respect to the cooling transition is $43\,\rm{MHz}$. This allows us to address atoms with velocity classes in the range of [$0,-45\,\rm{m/s}$] that would otherwise not contribute to the atom flux. The use of the pushing beam is found to improve the flux by a factor of $12\pm2$ as illustrated in Fig.~\ref{fig:2dmotmodule}(b). Varying the dispenser current offers a simple way to control the rubidium vapor pressure, which strongly influences the atomic flux and the MOT loading rate (see
Fig.~\ref{fig:2dmotmodule}(b)). A maximum 3D-MOT loading rate of  $(3.9 \pm 0.5) \times 10^{9}\,\mathrm{s^{-1}}$ is obtained for a dispenser current of 3.2\,A~\footnote{We use ALVASOURCES from Alvatec, which are chromate-free metal vapor sources of the type AS-3-Rb87(98\%)-20-F}. For higher currents background gas collisions become important such that the flux decreases again~\cite{dieckmann1998,schoser2002,catani2006, chaudhuri2006}. Under optimum conditions we trap $3.5 \times 10^{8}$ atoms in $200\,\rm{ms}$. Further details about the design and construction of the 2D-MOT can be found in~\cite{gotz2012}.

\subsection{B. Optimized three beam optical dipole trap}

The optical dipole trap (ODT) is designed as a horizontally oriented three beam trap and spatially overlapped with the 3D-MOT using a geometry which provides a large trapping volume and tight confinement. To fulfill these seemingly incompatible requirements, we decouple the loading from the evaporation process using a separate reservoir and dimple trap~\cite{jacob2011,clement2009,weber2002}. The large reservoir trap consists of two weakly-focused beams crossing at a small angle, while the dimple trap is tightly focused to enhance the thermalization rate. The ODT is generated from a $50\,\rm{W}$ single frequency fiber amplifier laser with a wavelength of at $1064\,\rm{nm}$ (approximately $22\,\rm{W}$ at the position of the atoms).

\begin{figure}
  \begin{center}
      \includegraphics[width=0.95\columnwidth]{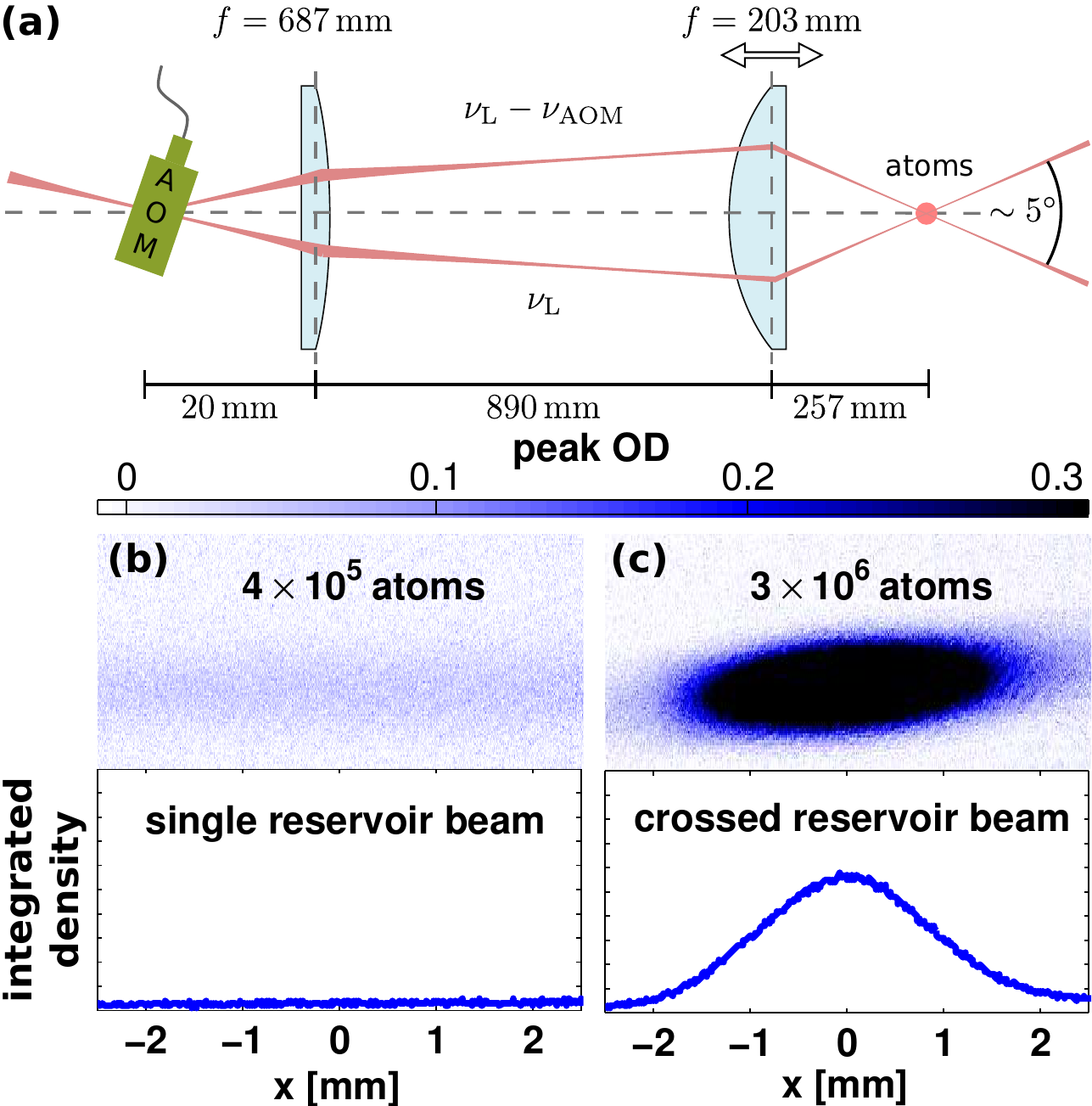}
		 \caption{(Color online) \textbf{Atom trapping in a crossed 1064\,nm reservoir trap.} 
		 (a) Schematic of the reservoir trap. An adjustable (double arrow) imaging system with a
		 $\approx0.3$ magnification is used to image the diffraction pattern of an acousto-optic
		 modulator (AOM) onto the atom cloud. 
		 (b) \& (c) show absorption images of a single beam (AOM off) and crossed (AOM on) reservoir 
		 trap, respectively after $2\,\rm{ms}$ of free expansion. The lower panels show the according 
		 integrated density distributions. Using the same all over laser beam power ($21.5\,\rm{W}$) in
		 both scenarios, the crossed beam trap confines almost one order of magnitude more atoms.}
    \label{fig:dipoletrap}
  \end{center}
\end{figure}

The cigar shaped reservoir trap is realized by imaging the Bragg diffraction pattern of an acousto-optic modulator (AOM) onto the atoms (see Fig.~\ref{fig:dipoletrap}(a)).
In this way we produce two horizontally polarized Gaussian laser beams of $120\,\rm{\mu m}$ waist crossing at an angle of $5.4^\circ$, and frequency shifted by the acousto-optical modulator (AOM) frequency $\nu_{\rm{AOM}}$
(typically $80\,\rm{MHz}$) which prevents laser beam interference. Imaging the AOM not only simplifies the crossing alignment but also allows for dynamical power balancing of the two intersecting reservoir arms through the radio frequency power. For a typical laser power of $10.7\,\rm{W}$ in each reservoir beam the trap depth is $U_{0,\rm{res}} \approx k_B\times145\,\rm{\mu K}$. Most importantly the crossed geometry leads to a significant enhancement of confinement along the symmetry axis of the reservoir. A dramatic consequence of this is illustrated in Fig.~\ref{fig:dipoletrap}(b)\,\&\,(c), where we compare the performance of a single beam reservoir to a crossed beam reservoir, both with the same trap depth and the same radial confinement. As compared to the single beam trap, we observe an increase in atom number by approximately one order of magnitude: from $4\times10^5$ to $3\times10^6$. This approach seems to be comparable to or even more efficient than the loading improvement using hybrid magnetic and optical traps~\cite{zaiser2011}. 

The dimple beam is a third independently controllable, vertically polarized laser beam focused to a waist of $23\,\rm{\mu m}$, crossing the reservoir axis under an angle of $45^\circ$, thereby creating an anisotropic ellipsoidal trapping potential. With a typical dimple laser power of $2\,\rm{W}$, we increase the depth of the potential by $U_{0,\rm{dimple}} \approx k_B\times350\,\rm{\mu K}$ such that a major fraction of the atoms initially collected in the reservoir are transferred into the dimple trap. The reservoir beams also serve to increase the confinement along the weak axis of the dimple trap.

\subsection{C. Preparation of dense atomic samples}

\emph{Experimental loading cycle ---}
Our experimental loading cycle is optimized to maximize the atom number in the reservoir trap and usually begins with a $550\,\rm{ms}$ loading phase at a dispenser current of 2.3\,A.  We typically load $3\times10^8$ atoms in the 3D-MOT while the three beam ODT is already switched on. The optimized parameters are: $125\,\rm{mW}$ total cooling power distributed over six 20\,mm diameter MOT beams, a red detuning from the cooling transition of $19\,\rm{MHz}$, and a magnetic field gradient of $10.5\,\rm{G/cm}$. The 2D-MOT is then switched off and the 3D-MOT is slightly compressed within $200\,\rm{ms}$ by increasing the magnetic field gradient to $14\,\rm{G/cm}$ and by ramping the cooler red  detuning to $25\,\rm{MHz}$. To further boost the density of the atomic cloud and hence optimize the loading of the ODT~\cite{kuppens2000}, we use a  $150\,\rm{ms}$ temporal dark MOT phase~\cite{townsend1996}. During this phase, the red detuning of the cooling light is ramped to $48\,\rm{MHz}$ and the total cooling power is reduced to $11\,\rm{mW}$. At the same time the power of the repumping laser is reduced from $140\,\rm{\mu W}$ to $1\,\rm{\mu W}$, resulting in a sizable population in the lower hyperfine ground state $|5S_{1/2},F = 1\rangle$. To minimize atom loss due to hyperfine changing collisions in the ODT, we pump all atoms into the $|5S_{1/2},F = 1\rangle$ manifold by turning off the repumping laser $2\,\rm{ms}$ before the end of the dark MOT phase. At this point, about $5\times10^6$ atoms have been transferred into the reservoir trap at a temperature of approximately $50\,\rm{\mu K}$. This is the ideal starting point for the evaporative cooling phase.

\begin{figure}
  \begin{center}
      \includegraphics[width=\columnwidth]{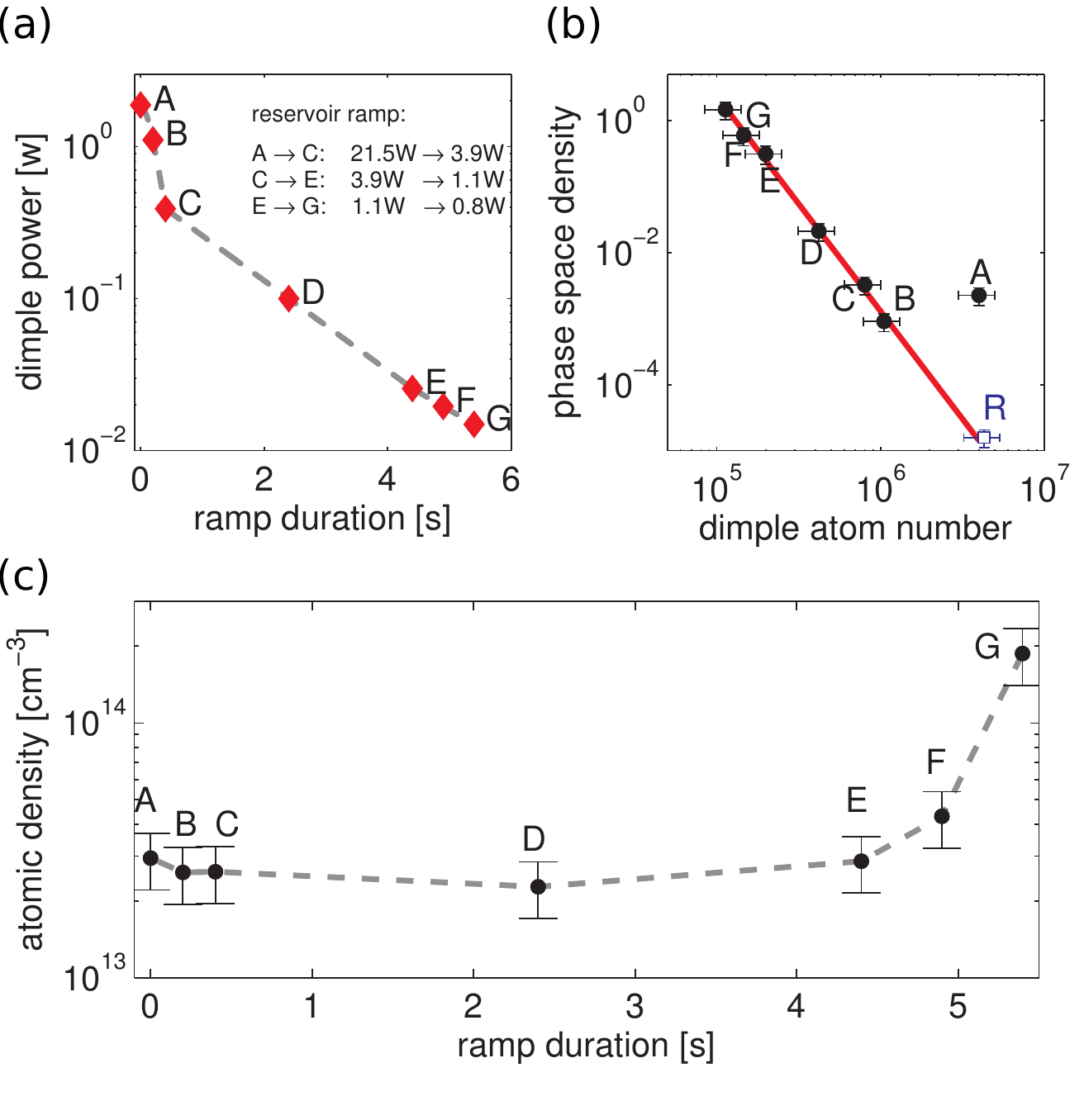}
      \caption{(Color online) \textbf{Optimized evaporative cooling ramp.} (a) Adiabatic lowering
	 of the dimple trap power. The evaporative cooling ramp is characterized at the points
	 'A-G'. The inset shows the according reservoir ramp. (b) Inferred phase-space density
	 $\varrho_0$ versus dimple atom number. The solid line is a power law fit to point 'B-G'
	 amounting to a cooling efficiency of $\gamma_{ev}=3.3\pm 0.3$. The open square `R' corresponds
	 to the exclusive use of the reservoir trap. (c) Shows the evolution of the atomic density
	 (in situ) in the combined trap during the evaporation ramp. The dashed lines serve as guide
	 to the eye.}
    \label{fig:evaporation_ramp}
  \end{center}
\end{figure}

\emph{Characterization of the evaporation ramp ---}
After $40\,\rm{ms}$ of free evaporation at constant trap depth we start a forced evaporation phase by adiabatically lowering the depth of the reservoir and the dimple trap. We found that active intensity stabilization of the dipole trap laser system during the evaporation ramp is not necessary. As the re-thermalization rate decreases during evaporation due to the reduced confinement, it is favorable to lower the trap depth at decreasing rate $dU_0/dt$ to keep the evaporation efficiency approximately
constant~\cite{ohara2001}. We achieve this with a sequence of three exponential ramps (A\,$\rightarrow$\,C, C\,$\rightarrow$\,E, E\,$\rightarrow$\,G) with durations of 0.4\,s, 4\,s, and
1\,s respectively (see Fig.~\ref{fig:evaporation_ramp}(a)).
The time constants for each ramp were experimentally optimized to obtain the highest gain in phase space density $\varrho_0$. Optimum results are found for a dimple power reduction by two orders of magnitude, while the reservoir power is only reduced by a factor of $\approx 25$.  The presence of the reservoir beams maintain strong confinement along the weak axis of the dimple throughout the evaporation process.

To characterize the evaporation ramp we first investigate the temperature decrease as a function of the number of atoms remaining in the dimple trap. We find that the temperature decreases by a factor of $\approx\,100$ while the atom loss only amounts to a factor of $\approx\,10$. The measured temperature rate of change is $\dot{T}/T=(2.3\pm0.3)\,\dot{N}/N$. Assuming harmonic confinement we calculate the atomic density during the evaporation sequence,
		\begin{equation}
			\label{eq:atomic_density}
				\rho_0=N\overline{\omega}^3 (m \lambda_{\rm{dB}}/h)^{3} \,,
		\end{equation}
where $\overline{\omega}$ is the geometrical mean of the calculated trapping frequencies and $\lambda_{\rm{dB}}$ is the thermal de Broglie wavelength. With the temperature $T$ and particle number $N$ it is possible to deduce the phase space density
		\begin{equation}
			\label{eq:phase_space_density}
				\varrho_0 = \rho_0 \lambda_{\rm{dB}}^3 \propto N T^{-3}\,. 
		\end{equation}

The black data points (A-G) in Fig.~\ref{fig:evaporation_ramp}\,(b) show $\varrho_0$ as a function of $N$. As compared to the exclusive use of the reservoir (blue square (R)), we win approximately two orders of magnitude in phase space density when confining the atoms in the three-beam trap. For later evaporation times (B-G), we observe an increase in phase space density with decreasing atom number, with a constant cooling efficiency
		\begin{equation}
			\label{eq:evaporation_efficiency}
				\gamma_{evap}=-\frac{\rm{ln}[\varrho_0/\varrho_0']}{\rm{ln}[N/N']} = 3.3\pm 0.3 \,,
		\end{equation}
as shown by the solid line in Fig.~\ref{fig:evaporation_ramp}\,(b). This compares favorably with other recent experiments: $\gamma_{evap} = 3.5 $~\cite{Lauber2011}, $\gamma_{evap} = 2.8\pm0.5$~\cite{clement2009}. Furthermore, we estimate the truncation parameter during evaporation $U_0/k_BT = 5.8$, which is typical for all-optical evaporation schemes. The critical phase space density is reached with $N\approx 1\times10^5$ atoms in the dipole trap. Figure ~\ref{fig:evaporation_ramp}\,(c) shows the evolution of the atomic density (in situ) during the evaporation ramp. The three-beam trap allows us to maintain a high density throughout the evaporation, since the confinement along the symmetry axis of the trap is almost kept constant via a moderate reduction of the reservoir power. After evaporation the trap can be switched off and by varying the expansion time the atomic density is then varied over several orders of magnitude.

\begin{figure}
  \begin{center}
      \includegraphics[width=\columnwidth]{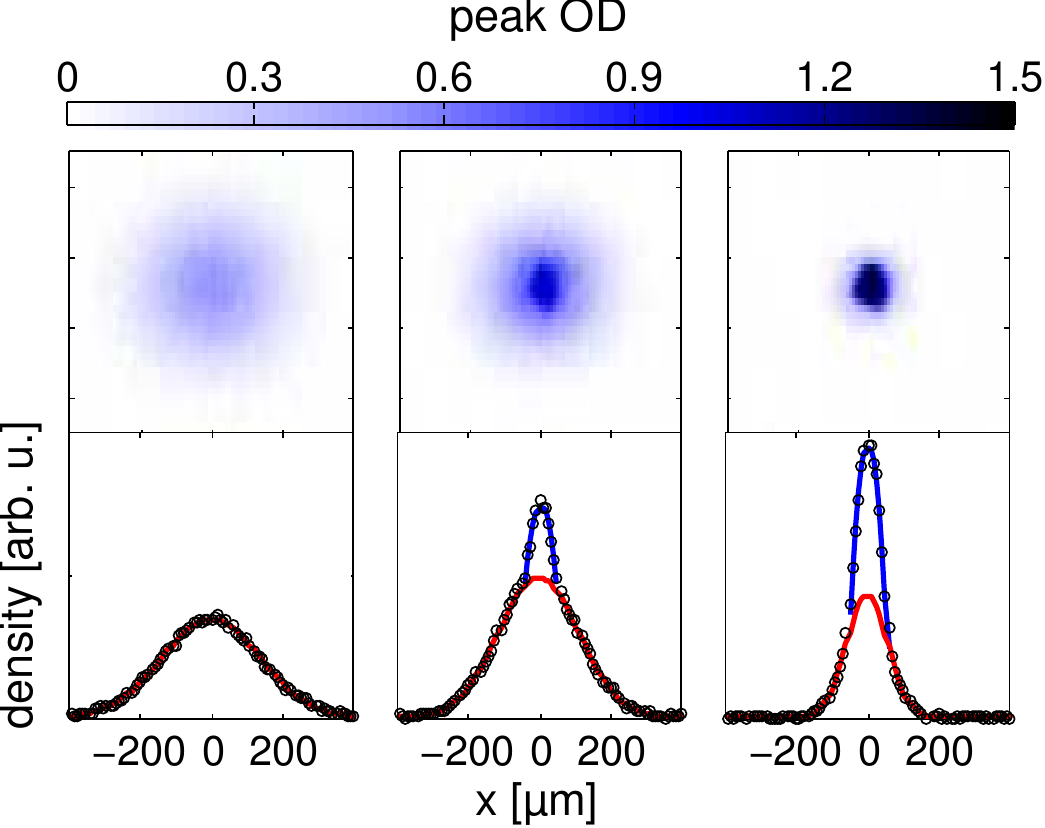}
      \caption{(Color online) \textbf{BEC-transition.} Upper row: Absorption images of a thermal
	    gas and two partially condensed gases, taken for various final ODT powers and after a 
        time-of-flight of 22\,ms. Lower row: Corresponding integrated atomic density distributions
	    (points). The second and third profile clearly show a bimodal structure evidencing the
	    onset of Bose-Einstein condensation. The red lines represent Gaussian fits to the
	    thermal fraction of the gas, whereas the blue lines are Thomas-Fermi profiles
	    quantifying the condensed fraction.}
      \label{fig:BEC}
    \label{fig:BEC_transition}
  \end{center}
\end{figure}

		\begin{figure*}
		  \begin{center}
		      \includegraphics[width=0.9\textwidth]{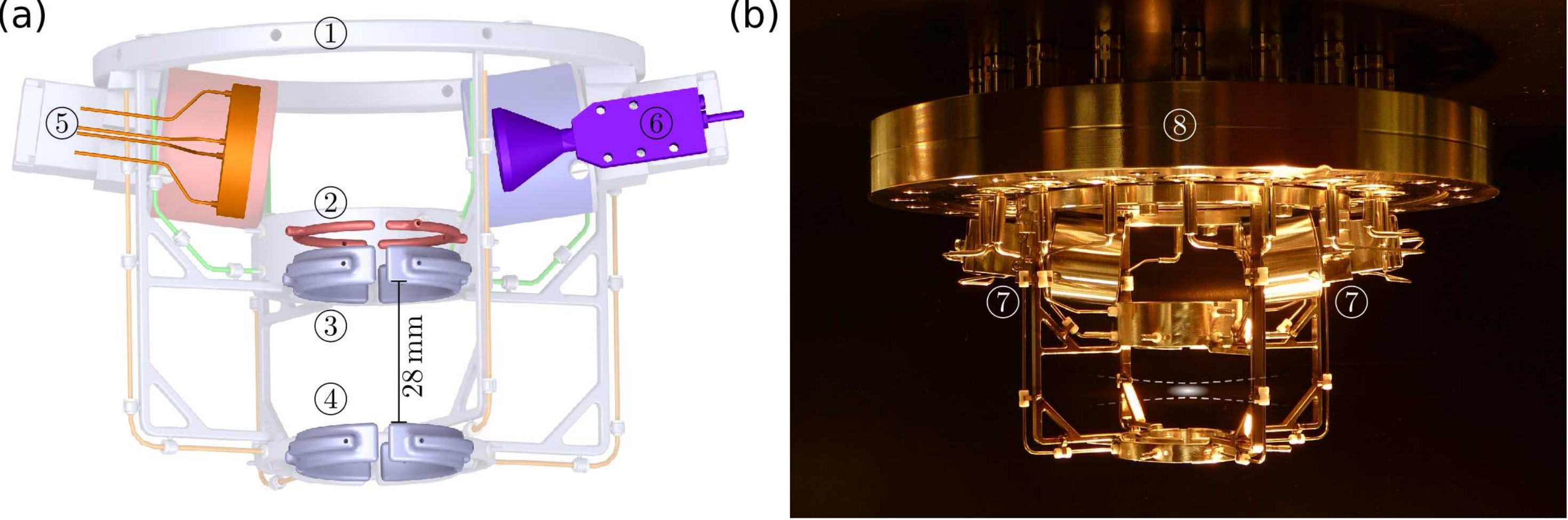}
				\caption[Electrode assembly for electric field control and  for field ionization of 	
				Rydberg atoms]{(Color online) \textbf{Electrode assembly for electric field control and  
				for field ionization of Rydberg atoms.}
		      (a) Schematic drawing of the electrode assembly for electric field control and  for field 
		      ionization of Rydberg atoms. A grounded supporting structure \circled{1} holds two 
		      deflection rings \circled{2} as well as four upper \circled{3} and four lower \circled{4} 
		      field plates in place, that are used to control the electric field landscape at the 
		      position of the atoms. Charged particle detection is possible with either a micro-channel 
		      plate detector (MCP)~\circled{5} or a channel electron multiplier detector (CEM) 
		      \circled{6}. 
		      (b) Photograph of the stainless steel structure which also shows the Faraday cages 
		      \circled{7} that surround the detectors and the vacuum flange \circled{8} with 18 high 
		      voltage connections. The position of the atom cloud and the orientation of the dipole trap
		      are sketched on the photograph.}
		    \label{fig:3_detector}
		  \end{center}
		\end{figure*}
\emph{BEC-transition ---}
To produce a Bose-Einstein condensate, the dimple trap depth is decreased further. At a trap depth
of approximately 700\,nK, the gas crosses the transition to a BEC. The expected critical temperature with $N \approx1 \times10^5$ is $T_c = 260\,\rm{nK}$. Reducing the trap depth further leads to an almost pure condensate with $\approx2 \times 10^4$ atoms. By Stern-Gerlach separation we measure the relative population of the three magnetic sub-levels to be 0.5, 0.3, and 0.2 in the $|m_F= -1\rangle$, $|m_F= 0\rangle$, and $|m_F= +1\rangle$ magnetic states of the $|5S_{1/2}, F= 1\rangle$ manifold, respectively. Using slower evaporation ramps, the best number achieved was $\approx6 \times 10^4$ atoms
in a 70\,\% pure condensate. The shortest \emph{total} cycle time which still produced a BEC was around 4.5\,s.

\subsection{D. Rydberg atom detection}

\emph{Rydberg detector ---} To manipulate and detect the Rydberg atoms we use a specially designed stainless steel electrode assembly (shown in Fig.~\ref{fig:3_detector}\,(a)~\&~(b)). It is designed in such a way that it does not obscure the optical access to the atom cloud, such that optical probing of Rydberg states and optical lattices can easily be implemented. The electrode structure can be used to null electric fields, to apply offset fields and field gradients, or to ionize Rydberg atoms (with fields up to $\gtrsim 1.2\,\rm{kV/cm}$) and guide the resulting ions to a charged particle detector.

The electrode structure is attached to the science chamber with a DN160CF to DN63CF reducing flange with 18 welded SHV connections (\circled{8}) as shown in Fig.~\ref{fig:3_detector}\,(b). An electrically grounded supporting structure (\circled{1}) holds in total ten individually addressable electrodes in place, which  are used to control the electric field landscape at the position of the atoms. The electrodes are divided into two deflection half rings (\circled{2}) and eight field plates as illustrated in Fig.~\ref{fig:3_detector}\,(a). The field plates are situated at the extremes of a virtual cylindrical volume of height $44\,\rm{mm}$ and diameter $34\,\rm{mm}$ and are arranged such that four field plates are situated `above' (\circled{3}) and four field plates are situated `beneath' (\circled{4}) the atoms. Each field plate resembles a quarter cylinder segment with an inner radius of $15\,\rm{mm}$, a height of $8\,\rm{mm}$ and a wall thickness of $2\,\rm{mm}$. 

\begin{figure}
  \begin{center}
      \includegraphics[width=0.9\columnwidth]{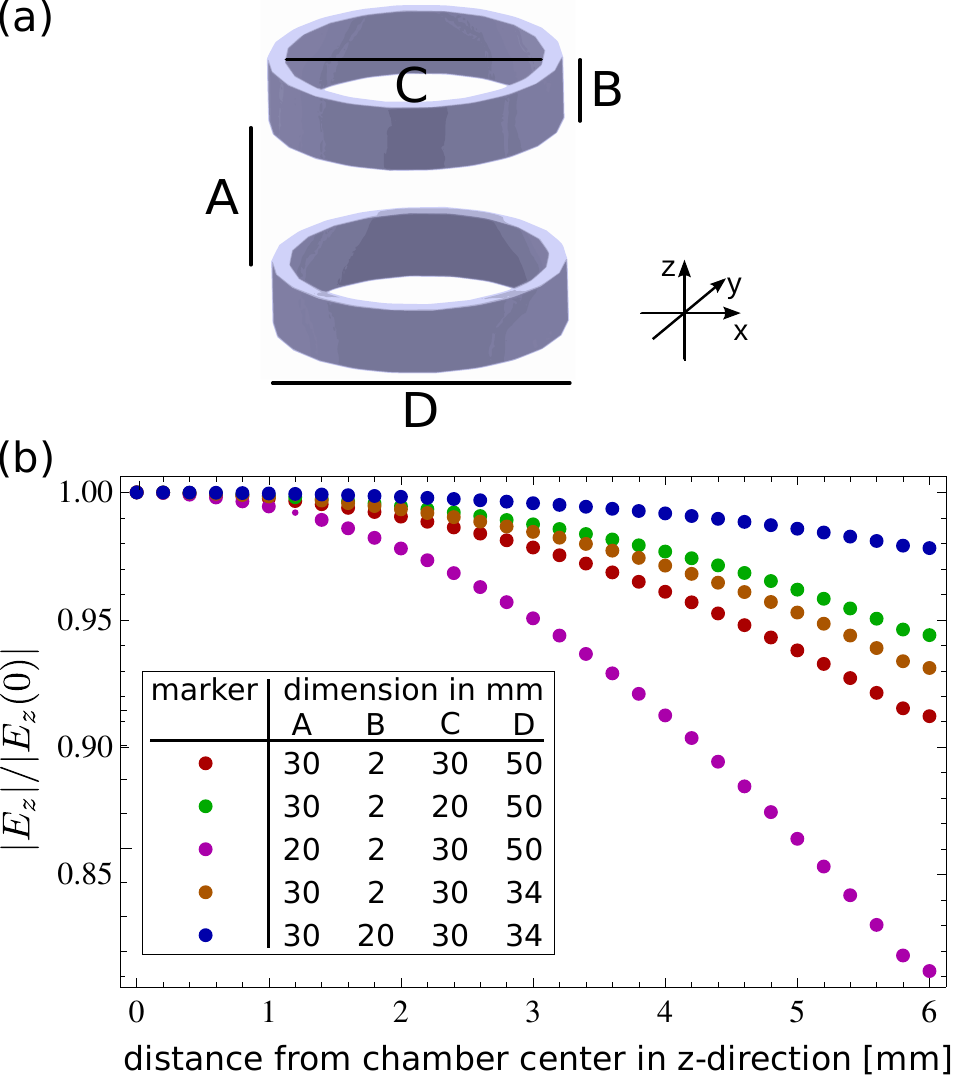}
      \caption{(Color online) \textbf{Optimization of the field plate geometry.} 
      (a) Basic field plate geometry with A $\equiv$ separation, B $\equiv$ thickness, C 
      $\equiv$ inner diameter, and D $\equiv$ outer diameter of the field plates. 
      (b) The normalized modulus of the z-component of the electric field is plotted as a function of
      the radial distance from the center of the chamber $(x=y=z=0)$. The optimization is shown for five 
      different parameter sets, which are specified in the inset of (b).}
    \label{fig:detector_optimization}
  \end{center}
\end{figure}

The shape and separation of the field plates and deflection rings were optimized with extensive SIMION-simulations to produce the lowest electric field inhomogeneities at the position of the atoms. 
An example of such a simulation based optimization is shown in Fig.~\ref{fig:detector_optimization}. Here we approximated the final shape of the eight field plates by two ring-shaped electrodes as shown in Fig.~\ref{fig:detector_optimization}\,(a). In the SIMION simulations we varied the geometry (with A $\equiv$ separation, B $\equiv$ thickness, C $\equiv$ inner diameter, and D $\equiv$ outer diameter) of the `test field plates' and computed the normalized electric field vector components. Figure~\ref{fig:detector_optimization}\,(b) shows the normalized modulus of the z-component of the electric field as a function of the radial distance from the center of the chamber for five different geometries. Similar results were obtained for the radial field components. In general, we observe that a prolate geometry results in smaller electric field inhomogeneities along the z-axis than an oblate geometry. Accounting for the grounded vacuum chamber, the exact course of the electrical wiring and the true shape of the field plates of the final geometry of the electrode structure depicted in Fig.~\ref{fig:3_detector}, the simulated field deviations only amount to $<0.5\,\%$ over a volume of 1\,mm$^3$.

We use bipolar high voltage switches with an effective rise time of $\approx300\,\rm{ns}$ to
dynamically control the electric fields and to perform field ionization of Rydberg atoms. We realize
this by applying large state-dependent ionization fields. Half-ring-shaped deflection electrodes (\circled{2}) are used to guide the produced ions on a curved trajectory to either a micro-channel plate detector (MCP; \circled{5}) or a channel electron multiplier detector (CEM; \circled{6}) as shown in Fig.~\ref{fig:3_detector}\,(a). The CEM and the MCP  both have a circular detection area of $165\,\rm{mm}^2$. Contrary to the design described in~\cite{low2012,Sassmannshausen2013}, the two detectors are mirror-symmetrically installed above the upper field plates to maintain optical access in the horizontal plane of the experiment. To minimize the stray fields produced by the $\approx2\,\rm{kV}$ bias voltages, both detectors are mounted inside Faraday cages (see \circled{7} in photograph in Fig.~\ref{fig:3_detector}\,(b)). As the detector has no direct line of sight to the ion's `place of birth' we obtain absolutely background-free detection signals with single-ion resolution.  From the measured ion-pulse-height-distribution we infer a counting efficiency of 0.9, limited by the electrical noise. Combined with the MCP detection efficiency ($\approx 0.6$), the angular responsivity ($\approx 0.9$) and the guiding efficiency ($\approx 0.9$) this gives an estimate for the overall efficiency of $\eta\approx 0.4$.

\begin{figure}
  \begin{center}
      \includegraphics[width=0.9\columnwidth]{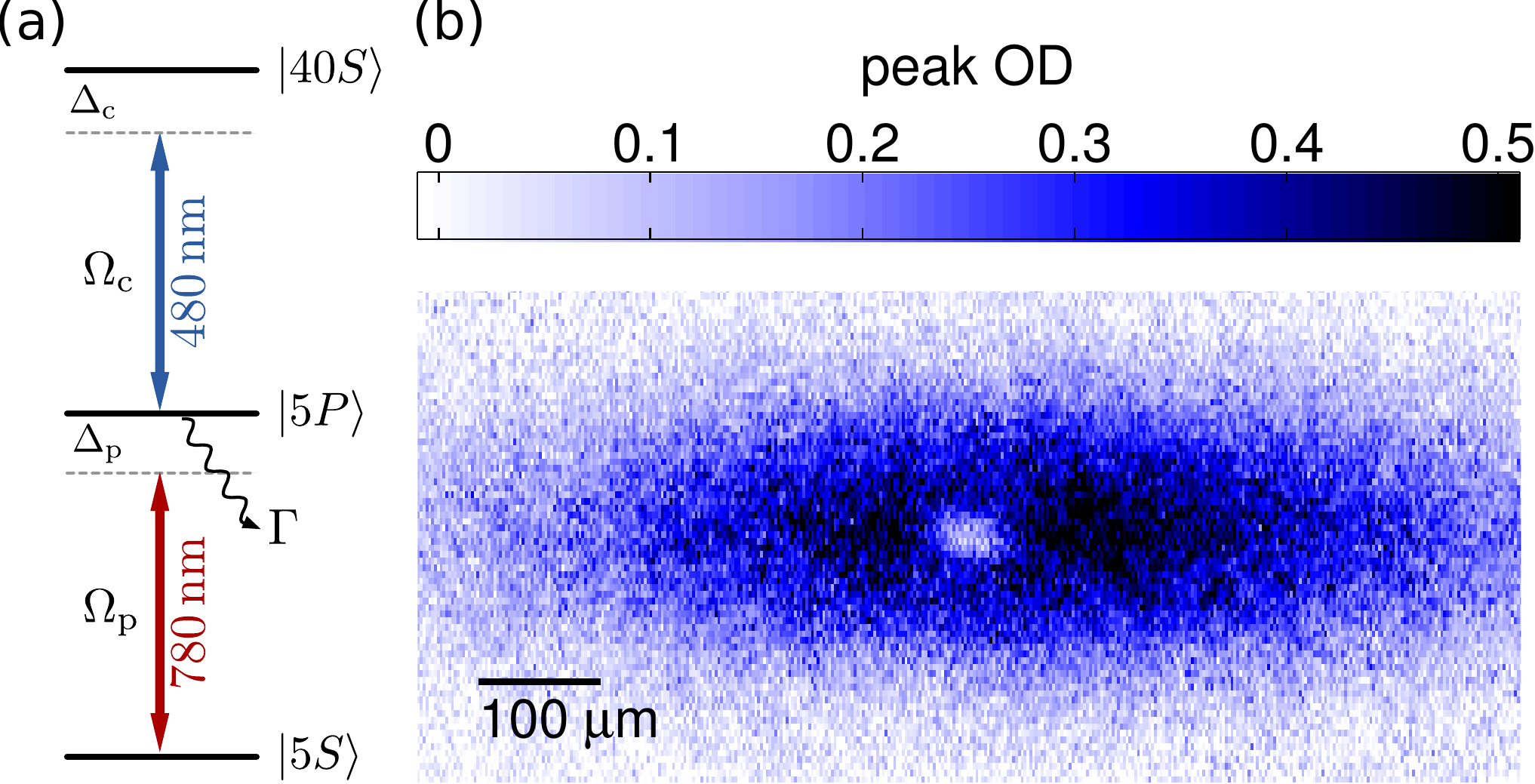}
      \caption{(Color online) \textbf{Optical probing of Rydberg states using electromagnetically induced 
      transparency.} 
      (a) Energy level diagram with $\Omega_{\rm{p}}$ ($\Omega_{\rm{c}}$) the Rabi frequency and
      $\Delta_{\rm{p}}$ ($\Delta_{\rm{c}}$) the laser detuning of the 780\,nm probe (480\,nm coupling)
      laser, respectively. $\Gamma$ is the spontaneous decay rates of the intermediate $\ket{5P}$ 
      state. 
      (b) Absorption image of the atomic cloud for $\Delta_{\rm{p}} = \Delta_{\rm{c}}= 0$, taken after an free expansion of $600\,\rm{\mu s}$. The induced 
      transparency inside the excitation volume is clearly visible in the center of the absorption
      image.}      
      \label{fig:EIT}
  \end{center}
\end{figure}

\emph{Optical probing of Rydberg states ---}
Another possibility to probe Rydberg state properties is the use of Rydberg electromagnetically induced transparency (EIT) as a means for spatially resolved optical detection with a CCD camera~\cite{Tauschinsky2010,Hattermann2012,Hofmann2013}. EIT is a quantum interference effect in which an optical control field renders an otherwise absorbing gas transparent~\cite{Fleischhauer2005}. The basic principle involves a three-level system composed of two long-lived states $|1\rangle$ and $|3\rangle$ that are coupled via a short-lived state $|2\rangle$ which spontaneously decays with rate $\Gamma$. In Rydberg state EIT the system is a ladder system with $|3\rangle$ being a Rydberg state. A strong laser coupling $\Omega_c$ results in an Autler-Townes doublet of dressed states, such that the transition amplitudes for the $|1\rangle \rightarrow |2\rangle$ resonance interfere destructively. This gives rise to a narrow transparency resonance which is mapped onto the light field and which can therefore be detected on a CCD camera. 

In the experiments detailed here, we chose the three-level ladder system depicted in Fig.~\ref{fig:EIT}\, (a), with: $|1\rangle \equiv |5S_{1/2},F=2\rangle$, $|2\rangle \equiv |5P_{3/2},F=3\rangle$, and 
$|3\rangle \equiv |40S_{1/2}\rangle$. For simplicity we moreover restrict ourselves to a regime of experimental parameters in which Rydberg--Rydberg interaction effects on the EIT can be neglected~\cite{Pritchard2010,Schempp2010,sevincli2011}. We prepare a cigar shaped Gaussian cloud ($\sigma_{\rm{radial}} = 75\pm3\,\rm{\mu m}$, $\sigma_{\rm{axial}} = 280\pm10\,\rm{\mu m}$) with $N\approx2.4\times10^5$ atoms and a peak atomic density $\rho = 2.2\times10^9\,\rm{cm}^{-3}$. The atoms are subsequently excited to $|40S_{1/2}\rangle$-Rydberg state, which does not exhibit significant Rydberg--Rydberg interactions at these densities. The Rydberg state is addressed via resonant two photon excitation, realized by a collimated (waist $= 15\,mm$) $780\,\rm{nm}$ probe beam and a counter propagating focused (waist $=25\,\mu m$) $480\,\rm{nm}$ coupling beam. The  $\sigma^+$ polarized probe laser has a Rabi frequency $\Omega_p$ while the $\sigma^-$ polarized coupling laser has peak Rabi frequency $\Omega_c$. Both lasers are switched on for an excitation time of $t_{\rm{ex}}=100\,\rm{\mu s}$.  During this time we collect the probe light, which uniformly illuminates the cloud, on a CCD camera
for optical detection (see Fig.~\ref{fig:setup_total}). The imaging system has an optical resolution of $\approx 12~\mu m$ (Raleigh criterion). The fact that EIT allows us to map
the properties of the Rydberg state onto the light field is directly seen on the recorded absorption
images: on the two photon resonance we observe a local transparency due to the presence of the coupling
beam (see Fig.\,\ref{fig:EIT}\,(b)). From a detailed image analysis we infer that the latter causes a $\approx 75$\% reduction of the two level absorption in the excitation volume.

\section{III Emergence of correlations in strongly-interacting ultracold Rydberg gases }

For neutral atoms, Rydberg--Rydberg interactions possess the strongest inter-particle interaction, exceeding typical interactions in cold gases by several orders of magnitude~\cite{Saffman2010}. In these systems the laser excitation process itself has the potential to create strongly correlated many-body states. The hallmark is the Rydberg blockade~\cite{Comparat2010}, in which the interaction between Rydberg atoms inhibits the simultaneous laser excitation of multiple atoms within the blockade radius $R_{\rm{bl}}$. This radius is defined as the distance between two atoms at which the Rydberg interaction energy equals the excitation bandwidth $\Omega $:
	\begin{equation}
   			R_{\rm{bl}}=\sqrt[6]{\frac{|C_6|}{\hbar \Omega}}\, .
	\end{equation}	
The strong Rydberg--Rydberg interactions typically results in blockade radii on the scale of several micrometers. For instance the $\ket{55\,S_{1/2}}$ state has $ C_6 = 2\pi \hbar \times 50\,\rm{GHz\,\mu m}^6 $ and for $ \Omega \approx 2\pi \times 1\,$MHz this results in a blockade radius of $ R_{\rm{bl}} \approx 5\,\rm{\mu m} $. The blockade effect determines the excitation dynamics of an ultracold Rydberg gas. Since the excitation is coherently shared amongst all atoms inside each blockade sphere, collective excitation effects can arise~\cite{Dicke1954,Heidemann2007,Gaetan2009,Urban2009,Dudin2012b}. Bulk properties of Rydberg gases are likewise influenced by the blockade phenomenon as evidenced by a strong suppression of excitation~\cite{Tong2004,Singer2004,vogt2006,Vogt2007} or sub-Poissonian counting statistics~\cite{Reinhard2008,Viteau2012,Hofmann2013}, as will be discussed in Sec.~III\,B.

		\begin{figure}
		  \begin{center}
		      \includegraphics[width=0.95\columnwidth]{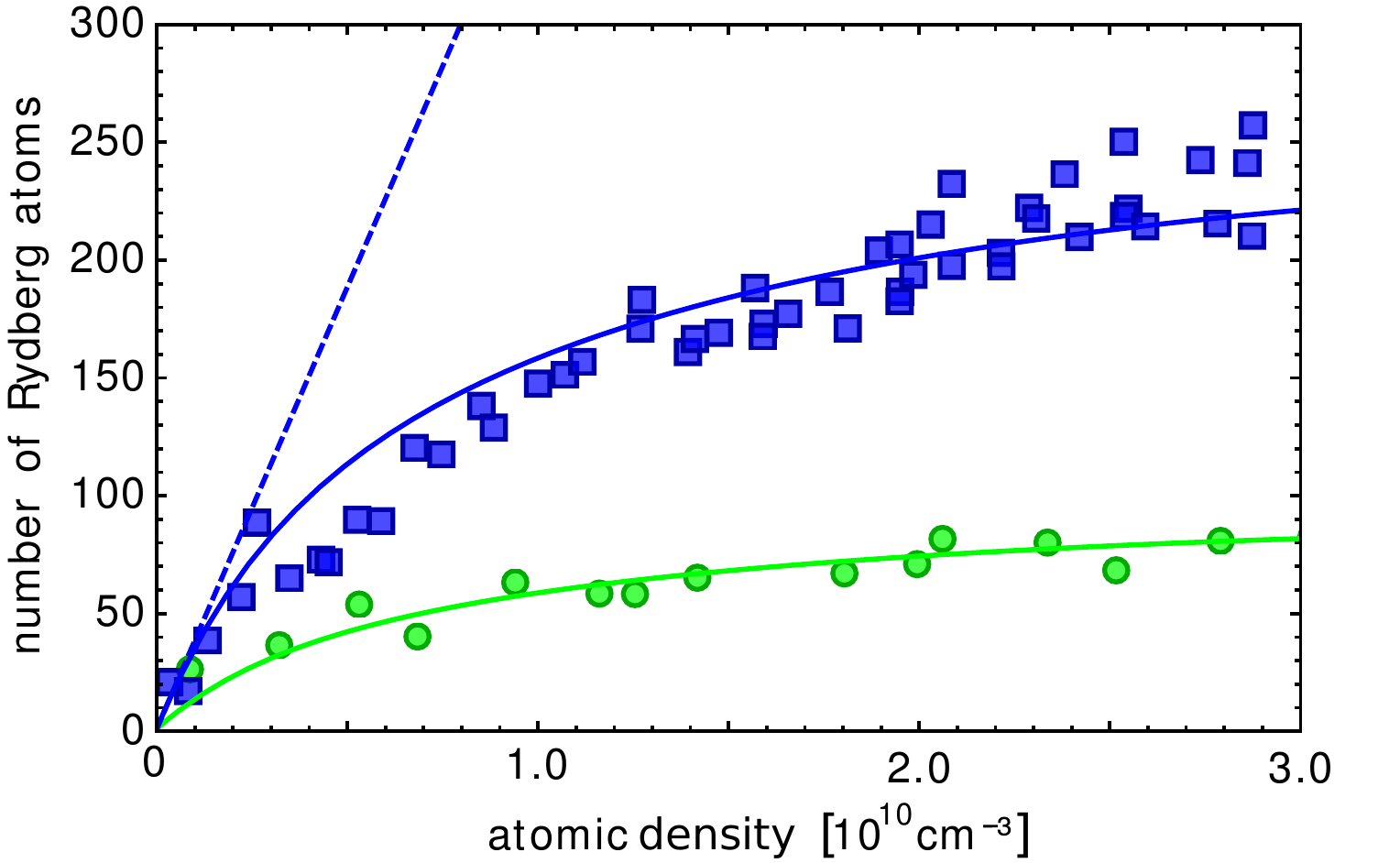} 
		      \caption{(Color online) \textbf{Measurement of interaction-induced Rydberg excitation 
		      suppression.} 
				The plot shows the $ \ket{55\,S} $ Rydberg atom number (corrected for the 
		      40\% detection efficiency) as a function of the atomic density in the excitation volume. 
		      It compares two sets of experiments done with different excitation volumes (see text). The 
		      solid lines are fits to the data using the stationary solution of a simple rate equation 
		      model. The dashed line indicates the Rydberg atom number for the neglect of interactions.}
		    \label{fig:2_blockade_measurement}
		  \end{center}
		\end{figure}

Figure~\ref{fig:2_blockade_measurement} shows a typical measurement of the suppressed excitation of $\ket{55\,S}$ Rydberg atoms as a function of atomic density. The latter was varied by controlling the population in the $\ket{5S_{1/2},F=2}$ state by optical pumping. The two measurements in Fig.~\ref{fig:2_blockade_measurement} were performed after different expansion times of the gas and the difference in the saturation Rydberg numbers reflects the relative sizes of the excitation volumes. The squares were obtained for an ellipsoidal excitation volume with Gaussian widths $\sigma_x \approx 40\,\rm{\mu m}$, $\sigma_y \approx 27\,\rm{\mu m}$, $\sigma_z \approx 27\,\rm{\mu m}$ after 200\,$\mu$s expansion time, whereas the circles were obtained for a volume defined by $\sigma_x \approx 40\,\rm{\mu m}$, $\sigma_y \approx 16\,\rm{\mu m}$, $\sigma_z \approx 16\,\rm{\mu m}$ and 50\,$\mu$s expansion time. The Rydberg atom number is corrected assuming a detection efficiency of $\eta=0.4$ and is in good agreement with the stationary solution of the rate equation model described in~\cite{SaintVincent2013}, using $R_{\rm{bl}} = 5\,\rm{\mu m} $ (solid lines). The dashed line indicates the anticipated Rydberg atom number neglecting interactions. 

Having discussed the preparation and detection of strongly-interacting Rydberg gases in detail, we now review two experiments in which many-body effects emerge in the strongly-interacting regime. Subsection A discusses the application of correlations between Rydberg atoms as a new route to strongly coupled ultracold plasmas. Subsection B then describes the observation of spatial and temporal correlations between Rydberg-interacting dark-state polaritons revealed by their sub-Poissonian counting statistics.

\subsection{A. Towards strongly coupled ultracold plasmas via Rydberg blockade}

\begin{figure*}
		  \begin{center}
		      \includegraphics[width=0.95\textwidth]{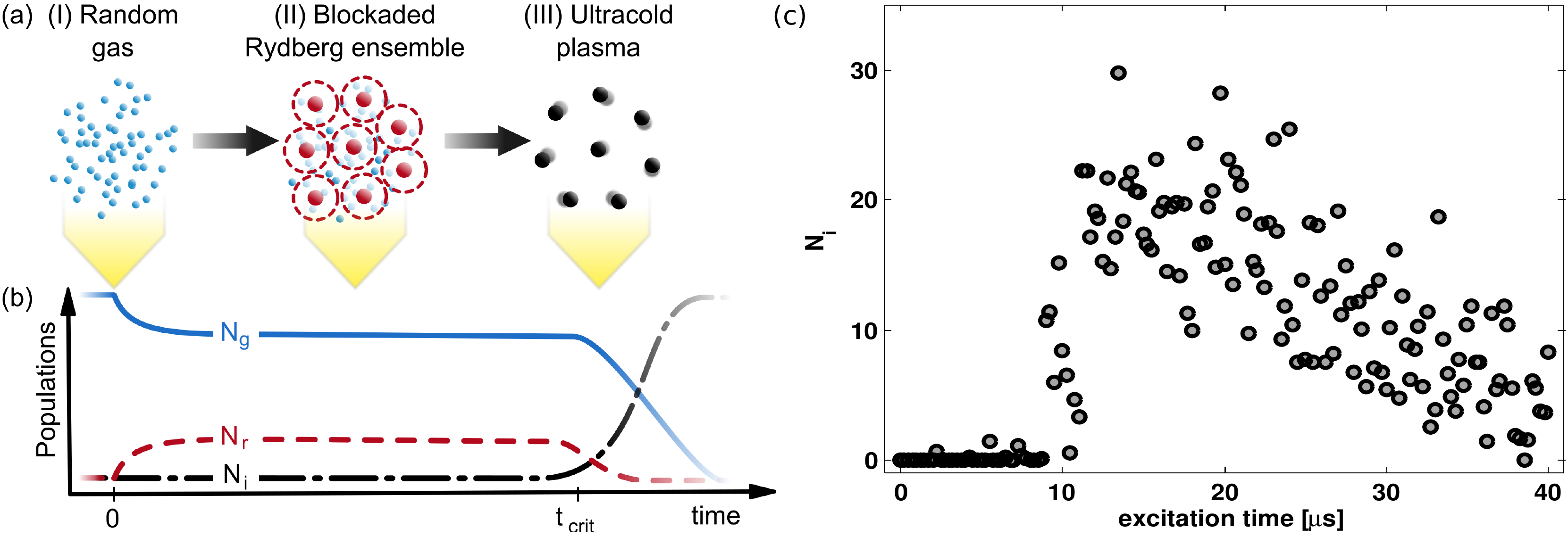}
		      \caption{(Color online) \textbf{Generation of an ultracold plasma from a Rydberg-blockaded
		       gas.} (a) An initially uncorrelated gas of ultracold atoms is prepared in an optical 
		       dipole trap (I). By a doubly resonant two-photon laser excitation some atoms are excited
		       to a Rydberg state. 
		       The strong Rydberg--Rydberg interactions prevent the excitation of close-by pairs, leading 
		       to spatial correlations (II). After a time $t_{\rm{crit}}$ the Rydberg gas is observed to 
		       spontaneously ionize into an ultracold plasma (III). 
		       (b) Qualitative dynamics of the involved populations: ground state atoms $N_g$ (solid 
		       line), Rydberg atoms $N_r$ (dashed line), and ions $N_i$ (dash-dotted line).
		       (c) Plasma ion signal taken at a ground state density of 
		       $\rho = 5\times 10^{10}\,\rm{cm^{-3}}$ in a volume of 
		       $\approx 1 \times 10^5\,\rm{\mu m}^3$. The rapid ionization avalanche starts at
		       $t_{\rm{crit}} \approx 10\,\rm{\mu s}$.
		       }
		    \label{fig:5_plasma_mechanism}
		  \end{center}
\end{figure*}

In dense samples, ionizing collisions between Rydberg atoms can produce an ion cloud which traps electrons produced during later collisions. Subsequently the trapped electrons rapidly collisionally ionize the remaining Rydberg atoms, resulting in an ensemble of unbound charged particles, referred to as an ultracold plasma~\cite{Robinson2000,Killian2007}. This spontaneous evolution of a Rydberg-interacting gas to an ultracold plasma has been observed in several experiments~\cite{Li2004,walz-flannigan2004,Morrison2008,SaintVincent2013}, providing a novel approach to study ultracold plasmas.
Ultracold plasmas constitute an ideal laboratory system to model exotic phases of matter such as laser-induced plasmas~\cite{Remington1999}, dense astrophysical plasmas~\cite{Horn1991} or even quark gluon plasmas~\cite{Shuryak2009}. Research in this field is fueled by the potential to access the so-called strongly coupled regime, in which the Coulomb interaction energy exceeds the kinetic energy of the ions, giving rise to many-body effects and strong spatial correlations between particles~\cite{Ichimaru1982,Killian2007b}. This regime is quantified by the Coulombic coupling parameter
	\begin{equation}
			\mathcal{G} = \frac{q^2}{4 \pi \epsilon_0 a k_B T} \gg 1\,.
		\label{eq:coupling_parameter}
	\end{equation}
Here, the charge and the temperature of the plasma constituents are given by $q $ and $T$, respectively, while $a = \sqrt[3]{3/(4 \pi \rho)} $ is the Wigner-Seitz radius, which characterizes the separation
between particles in a plasma of density $\rho$. 
To date ultracold plasmas with $\mathcal{G} \approx 0.1 - 2$ have been realized~\cite{Simien2004,Cummings2005}. However entering deep into the strongly coupled regime has remained out of reach despite the ultralow temperatures possible with laser cooling. Rapid temperature rises were observed during the plasma formation, which is partly attributed to the initially random distribution of the atoms~\cite{Kuzmin2002,Pohl2004,Bergeson2011}. This disorder-induced heating arises when the Coulomb interaction energy of the initial spatial distribution is converted into kinetic energy of the ions, posing one of the primary limits to the attainable coupling strengths in current experiments.

One solution to overcome disorder-induced-heating, is to produce the plasma from a Rydberg blockaded sample, in which the constituent particles are spatially ordered, with the probability to find close-by pairs strongly suppressed~\cite{SaintVincent2013,Bannasch2013}. This approach is illustrated in Fig.~\ref{fig:5_plasma_mechanism} and works as follows: (I) An initially randomly distributed gas of ultracold neutral atoms is excited to \textit{repulsively interacting} Rydberg states via a continuous and doubly resonant two-photon laser coupling. (II) Due to the Rydberg blockade, each Rydberg-excited atom blocks further excitations within the blockade radius leading to spatial correlations, resembling those of a strongly coupled plasma~\cite{Bannasch2013}. After a short time, the Rydberg density reaches steady state, while at later times Rydberg atoms start to decay through a combination of blackbody photoionization~\cite{Beterov2007} and ionizing collisions~\cite{Barbier1987,Kumar1999}, both leading to a successive increase of charged particles. (III) Once a critical number of ions has accumulated the resulting space charge starts to trap subsequently produced electrons. At this critical time $t_{\rm{crit}}$, rapid electron--Rydberg collisions trigger an ionization avalanche, which ultimately produces an ultracold plasma. Interestingly, the plasma forms rapidly compared to the motional timescales of the ions, since the avalanche is triggered by the fast moving electrons. Therefore the original density--density correlations of the Rydberg gas should be preserved, which is expected to significantly reduce the effects of disorder-induced heating~\cite{Murillo2001,Bannasch2013}. 

We have recently performed experiments along these lines, in which we observed the spontaneous ionization dynamics for Rydberg gases of different densities at temperatures around 
$230\,\rm{\mu K}$~\cite{SaintVincent2013}.
This was possible by independently measuring the number of ground state atoms, Rydberg atoms, and ions within the sample via the combination of absorption imaging and ionization detection. Evidence for the ionization avalanche deep in the blockade regime is given in Fig.~\ref{fig:5_plasma_mechanism}\,(c), which shows the number of plasma ions as a function of excitation time. For an atomic density of $\rho = 5\times 10^{10}\,\rm{cm^{-3}}$ we witness a sharp increase of plasma ions around $t_{crit} \approx 10\,\rm{\mu s}$, which is strongly density dependent. We observe that repulsive Rydberg--Rydberg interactions delay the ionization avalanche by several micro seconds. We found that the typical plasma formation time scales are around $2\,\rm{\mu s}$, which is short compared to the motional dynamics of the ions. During this time the ions only move by approximately 600\,nm, which is an order of magnitude smaller than the length scale imposed by the blockade radius $R_{\rm{bl}}$. This suggests that the initial correlations should be preserved in the plasma phase, putting the production of strongly coupled ultracold plasmas within reach. 

Recent molecular dynamics simulations for a gas of Sr atoms by G.~Bannasch \textit{et al.} predict that field ionization of a pre-structured Rydberg gas would suppress the effects of disorder-induced heating by more than an order of magnitude, giving rise to coupling parameters of up to $\mathcal{G} \approx 30 $, ~\cite{Bannasch2013}. Future studies of these systems will benefit from independent tuning of plasma parameters through appropriate choice of atomic densities, Rydberg quantum numbers and laser detunings. This would provide the means to reveal a proposed universal scaling between the coupling parameter $\mathcal{G}$ and the number of Rydberg blocked atoms~\cite{Bannasch2013} or to explore the phase diagram of ultracold plasmas including strongly coupled liquid phases and liquid-vapor phases~\cite{Shukla2011}.

\subsection{B. Sub-Poissonian counting statistics of strongly-interacting Rydberg-Polaritons}

\begin{figure*}
	\includegraphics[width=0.70\textwidth]{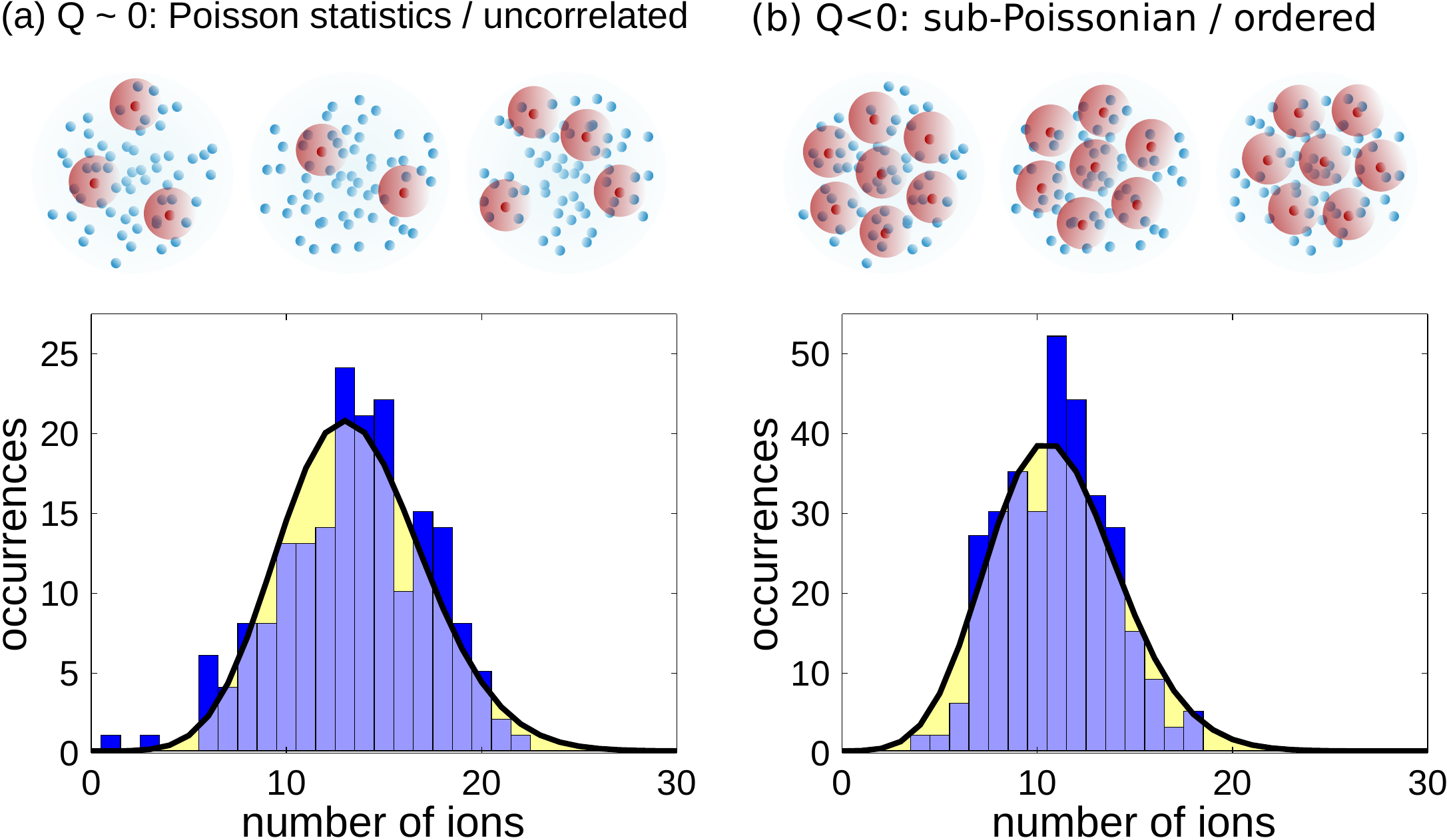}
	\caption{(Color online) \textbf{Counting statistics of Rydberg interacting dark-state polaritons.} 
	(a) For low polariton densities we anticipate uncorrelated polariton fluctuations and Poissonian 
	counting statistics with Mandel parameters $Q\approx 0$. The lower panel shows counting statistics 
	for 190 consecutive experiments in the low-density regime with $\left\langle N\right\rangle = 13.5 
	\pm 0.3$ and $Q = 0.05^{+0.11}_{-0.10}$.	
	(b) Contrary at high densities, we expect a strong suppression of the polariton number fluctuations
	reflected in sub-Poissonian counting statistics with $Q<0$. The lower panel shows counting 
	statistics for 320 consecutive experiments in the interacting regime yielding 
	$\left\langle N\right\rangle = 10.99 \pm 0.15$ and $Q = -0.31^{+0.04}_{-0.05}$. The thick curves in 
	both histograms show Possionian distributions with the corresponding means for comparison.}
	\label{fig:statistics}
\end{figure*}

For high densities the number of Rydberg excitations in the gas should be determined by the maximum number of blockade spheres that fit into the excitation volume. Therefore, the total number of Rydberg excitations saturates and is only subjected to small fluctuations between consecutive experimental runs. This can be seen in the statistical number distributions of Rydberg excitations which changes from Poissonian to sub-Poissonian character in the transition from the weakly-interacting to the strongly-interacting regime (see Fig.~\ref{fig:statistics}\,(a)~\&~(b)) \cite{Reinhard2008,Viteau2012,Hofmann2013}. This crossover can be quantified by the Mandel $Q$ parameter~\cite{Mandel1979}
\begin{equation}
 		\label{eq:mandel_q}
   			Q = \frac{\langle N_r^2\rangle - \langle N_r\rangle^2}{\langle 								N_r\rangle} -1\,,
	\end{equation}
where $N_r$ denotes the number of Rydberg atoms. Values of $Q>0$ correspond to super-Poissonian, $Q=0$ to Poissonian and $ Q<0$ to sub-Poissonian fluctuations. The latter corresponds to a narrowing of the Rydberg excitation number distribution. The extreme case, a number- or Fock-state is characterized by the total absence of fluctuations with $Q = -1$. Therefore the $Q$ parameter is an excellent measure for the effectiveness of the Rydberg blockade on the many-body system.

The Mandel $Q$ parameter can also be related to the spatial correlation function $ g^{(2)}(R) $ of the Rydberg atom distribution by
	\begin{equation}
 		\label{eq:q_correlation}
   			Q = \int  \rho_r \left[g^{(2)}(R) - 1 \right] d^3R\,, 
	\end{equation}
	with $\rho_r$ being the Rydberg density in the excitation volume~\cite{Wuster2010}. Even though (\ref{eq:q_correlation}) is only valid for homogeneous densities and isotropic interactions, it nonetheless underlines the fact that sub-Poissonian counting statistics are  intimately connected with the emergence of spatial correlations, reflected by $ g^{(2)}(R) $. Statistical methods for observing the transition to strongly-correlated or even crystalline states have recently been discussed in a number of theoretical papers~\cite{Breyel2012,Garttner2012b,Ates2012}. 
	
We recently reported experimental measurements of sub-Poissonian statistics of dark-state polaritons (DSPs) propagating in a gas of Rydberg-dressed atoms~\cite{Hofmann2013}. DSPs arise under the conditions of electromagnetically induced transparency~\cite{Fleischhauer2005}, in which a strong optical control field $\Omega_c$ renders an otherwise absorbing medium transparent for a weak probe field. At the few photon level, EIT can be described by the propagation of individual dark-state polaritons, which simultaneously possess the properties of both the photonic and the atomic degrees of freedom, which can be interchanged in a fully coherent and reversible way~\cite{Fleischhauer2000}.

The DSP wave function for a single probe photon can be written as
		\begin{equation}
		 	\label{eq:polariton}
				    \ket{D,1} = \cos\theta \ket{G,1} - \sin\theta \ket{R,0}\,.
		\end{equation}
where the collective quantum states involving $N$ atoms and the light field are 
		\begin{subequations}
			\label{eq:collective states}
			\begin{equation}
		 			\ket{G,1}  = \ket{g_1,\cdots , g_i, \cdots ,g_N}\ket{1}
			\end{equation}
			\begin{equation}
		 	\ket{R,0}  = \frac{1}{\sqrt{N}} \sum_{i} \ket{g_1,\cdots , r_i, \cdots ,g_N}\ket{0}\, .
			\end{equation}
		\end{subequations}
The mixing ratio $\tan(\theta)=\sqrt{\rho \sigma c \Gamma}/\Omega_{\rm{c}}$ determines the properties of the polariton depending on the atom density $\rho$, the resonant scattering cross-section $\sigma$, the speed of light in vacuum~$c$, the decay rate $\Gamma$ of the intermediate $\ket{5P_{3/2}}$ state, and the coupling Rabi frequency $\Omega_{c}$.

In our experiments, $|R\rangle=\ket{55S_{1/2}}$ and the DSPs are almost entirely matter-like with $\cos^2(\theta)\approx 10^{-4}-10^{-8}$. Therefore, measuring the Rydberg population serves as a projective measurement of the number of polaritons inside the cloud. This is achieved through single-atom sensitive detection of the Rydberg population which gives direct access to the matter part of the polariton wave function inside the atomic cloud. Rydberg--Rydberg interactions give rise to an effective interaction blockade for DSPs, which we observe as a transition from Poissonian to sub-Poissonian statistics for the polariton number, indicating the emergence of spatial and temporal correlations between dark-state polaritons. 

Figures~\ref{fig:statistics}\,(c)~\&~(d) show a comparison of the statistical distributions for low and high polariton densities corresponding to low and high atomic densities. For low atomic densities of $\rho \approx 2\times 10^{10}\,\rm{cm^{-3}}$, we obtain a distribution compatible with a Poissonian distribution and Mandel Q parameter of $Q = 0.05^{+0.11}_{-0.10}$ (see Fig.~\ref{fig:statistics}\,(c)). In contrast, for $\rho \approx 8\times 10^{11}\,\rm{cm^{-3}}$, we observe a substantial narrowing of the polariton number distribution, with $Q = -0.31^{+0.04}_{-0.05}$ (see Fig.~\ref{fig:statistics}\,(d)), where the experimental uncertainties have been estimated using the bootstrap method. The true distribution ($Q_{DSP}= Q/\eta$) is even narrower as one also has to take into account the ion detection efficiency $\eta \approx 0.4$~\cite{Ates2006,Reinhard2008,Viteau2012}. This indicates that the Rydberg--Rydberg interactions give rise to significant spatial correlations between DSPs. 

Effective polariton--polariton interactions mediated by the atomic admixture can lead to highly nonlinear~\cite{Pritchard2010,Petrosyan2011,Peyronel2012,Hofmann2013} and nonlocal optical effects~\cite{Sevinccli2011}, which allow the engineering of novel photonic many-body states. By detuning from the intermediate state, the dissipative character of the effective polariton--polariton interactions can be suppressed while maintaining the giant nonlinearities~\cite{Parigi2012}. This would give rise to elastic scattering of DSPs, providing means to implement deterministic photonic quantum logic gates~\cite{Shahmoon2011,Gorshkov2011}. As the nonlinearity can be controlled via the choice of the Rydberg state, repulsive, attractive and long-range interactions would become possible~\cite{Sevinccli2011}. This could pave the way to create and study new types of quantum gases such as dipolar Bose-Einstein condensates of dark-state polaritons~\cite{Fleischhauer2008,Nikoghosyan2012}. The ability to produce and coherently control the propagation of quantum fields using interacting dark-state polaritons will open up new applications including few-photon nonlinear optics~\cite{Pritchard2013}, nonclassical light sources~\cite{Honer2011,Dudin2012,Peyronel2012,Stanojevic2012,Maxwell2013} and photonic quantum logic gates~\cite{Friedler2005,Petrosyan2008}.

\section{IV Conclusion and Outlook}

In this paper we have presented a new experimental system for the investigation of many-body effects in dense ultracold gases of Rydberg atoms. The setup combines high atomic densities with excellent control of external fields and precise Rydberg atom detection. Fast experimental duty cycles are made possible using a high flux 2D-MOT followed by efficient evaporative cooling in an optimized three-beam optical dipole trap, which allows for the production of all-optical $^{87}$Rb BECs. We have described a specially designed electrode structure for electric field control and field ionization and detection of Rydberg atoms. This was then used to study the effects of correlations induced by Rydberg--Rydberg interactions in two vastly different physical regimes. 

The first study exploited Rydberg--Rydberg interactions as a means to imprint spatial correlations into an otherwise randomly arranged gas. Subsequent formation of an ultracold plasma was evidenced by a rapid avalanche ionization, suggesting a new route to the production of strongly coupled plasmas~\cite{SaintVincent2013}. The second study used measurements of the counting statistics to reveal temporal and spatial correlations between Rydberg-interacting dark-state polaritons. Here field ionization measurements served as a projective measurement of the polariton wave function inside the Rydberg medium~\cite{Hofmann2013}. 

In future experiments field ionization of Rydberg states, as described here, could be combined with optical probing of Rydberg states using absorption imaging under EIT conditions~\cite{Tauschinsky2010,Hattermann2012} (compare Fig.~\ref{fig:setup_total}). The combination of both techniques will provide information about Rydberg state populations and optical coherences with spatial resolution~\cite{tobepublished}, thereby giving dynamical access to macroscopic and microscopic properties of the system. 

Another future goal will be to explore strongly-correlated phases at the single particle level and with high spatial resolution. With this in mind we recently proposed a new imaging technique exploiting the large level shifts induced by a Rydberg impurity atom on a background gas of atoms, that is coupled to an auxiliary Rydberg state via an EIT resonance~\cite{Gunter2012}. Associated with this is a new characteristic radius within which the impurity--probe interactions destroy the transparency, thereby changing the optical properties for many atoms surrounding each impurity. 
This method should make it possible to image individual Rydberg atoms with high spatial and temporal resolution.  This technique is fundamentally nondestructive, thus allowing for time-resolved images of many-body dynamics and can be implemented in existing experiments. Numerous applications of this imaging technique can be foreseen, for example the possibility to directly image the formation (and melting) of crystalline states~\cite{Pohl2010,Schachenmayer2010,Vanbijnen2011}. 
Other studies could aim at spatially resolving individual dark-state polaritons propagating through the EIT medium and to directly reveal the nonlocal character of the optical nonlinearity~\cite{Sevinccli2011}. Ultimately, the ability to exploit Rydberg--Rydberg interactions to control both light and atomic fields, and to image the resulting new quantum phases will open a new chapter of Rydberg physics with many new and intriguing possibilities.

\section{Acknowledgements}
\acknowledgements{We acknowledge the tremendous support of the mechanical and electronic workshops and the construction office of the Physics Institute of the University of Heidelberg. In particular we want to thank J. Gerh\"auser for the mechanical design and S. Rabenecker for the assembly of the electrode structure, as well as K. Stumpf for the mechanical design of the 2D-MOT cage system. This work is supported in part by the Heidelberg Center for Quantum Dynamics and the Deutsche Forschungsgemeinschaft under WE2661/10.2. M.R.D.S.V. (grant number FP7-PEOPLE-2011-IEF-300870) and S.W. (grant number PERG08-GA-2010-277017) acknowledge support from the EU Marie-Curie program.}


\begin{thebibliography}{100}

\bibitem{Saffman2010}
Saffman, M., Walker, T.~G., and M\o{}lmer, K.
{\em Rev. Mod. Phys.} {\bf 82}, 2313--2363 (2010).

\bibitem{Comparat2010}
Comparat, D. and Pillet, P.
{\em J. Opt. Soc. Am. B} {\bf 27(6)}, A208--A232 (2010).

\bibitem{Pritchard2013}
Pritchard, J.~D., Weatherill, K.~J., and Adam, C.~S.
Non-linear optics using cold Rydberg atoms; Annual Review of Cold Atoms and
  Molecule chapter~8, pp. 301 -- 350
World Scientific, Singapore (2013).

\bibitem{Robicheaux2005}
Robicheaux, F. and Hern{\'a}ndez, J.
{\em Physical Review A} {\bf 72(6)}, 063403 (2005).

\bibitem{Weimer2008}
Weimer, H., L{\"o}w, R., Pfau, T., et al.
{\em Physical Review Letters} {\bf 101(25)}, 250601 (2008).

\bibitem{schwarzkopf2011}
Schwarzkopf, A., Sapiro, R.~E., and Raithel, G.
{\em Phys. Rev. Lett.} {\bf 107}, 103001 (2011).

\bibitem{Schauss2012}
Schau\ss\enspace, P., Cheneau, M., Endres, M., et al.
{\em Nature} {\bf 491}, 87 (2012).

\bibitem{Ates2012}
Ates, C. and Lesanovsky, I.
{\em Phys. Rev. A} {\bf 86}, 013408 (2012).

\bibitem{Petrosyan2013}
Petrosyan, D., H\"oning, M., and Fleischhauer, M.
{\em Phys. Rev. A} {\bf 87}, 053414 (2013).

\bibitem{Pohl2010}
Pohl, T., Demler, E., and Lukin, M.~D.
{\em Phys. Rev. Lett.} {\bf 104(4)}, 043002 (2010).

\bibitem{Schachenmayer2010}
Schachenmayer, J., Lesanovsky, I., Micheli, A., et al.
{\em New Journal of Physics} {\bf 12(10)}, 103044 (2010).

\bibitem{Vanbijnen2011}
{van Bijnen}, R. M.~W., Smit, S., {van Leeuwen}, K. A.~H., et al.
{\em Journal of Physics B: Atomic, Molecular and Optical Physics} {\bf 44(18)},
  184008 (2011).

\bibitem{santos2000}
Santos, L., Shlyapnikov, G.~V., Zoller, P., et al.
{\em Phys. Rev. Lett.} {\bf 85}, 1791 (2000).

\bibitem{pupillo2010}
Pupillo, G., Micheli, A., Boninsegni, M., et al.
{\em Phys. Rev. Lett.} {\bf 104}, 223002 (2010).

\bibitem{henkel2010}
Henkel, N., Nath, R., and Pohl, T.
{\em Phys. Rev. Lett.} {\bf 104}, 195302 (2010).

\bibitem{henkel2012}
Henkel, N., Cinti, F., Jain, P., et al.
{\em Phys. Rev. Lett.} {\bf 108}, 265301 (2012).

\bibitem{SaintVincent2013}
{Robert-de-Saint-Vincent}, M., Hofmann, C.~S., Schempp, H., et al.
{\em Phys. Rev. Lett.} {\bf 110}, 045004 (2013).

\bibitem{Bannasch2013}
Bannasch, G., Killian, T.~C., and Pohl, T.
{\em Phys. Rev. Lett.} {\bf 110}, 253003 (2013).

\bibitem{Dudin2012}
Dudin, Y.~O. and Kuzmich, A.
{\em Science} {\bf 336}, 887 (2012).

\bibitem{Peyronel2012}
Peyronel, T., Firstenberg, O., Liang, Q.~Y., et al.
{\em Nature} {\bf 488}, 57 (2012).

\bibitem{Maxwell2013}
Maxwell, D., Szwer, D.~J., Paredes-Barato, D., et al.
{\em Phys. Rev. Lett.} {\bf 110}, 103001 (2013).

\bibitem{Hofmann2013}
Hofmann, C.~S., G\"unter, G., Schempp, H., et al.
{\em Phys. Rev. Lett.} {\bf 110}, 203601 (2013).

\bibitem{Sevinccli2011}
Sevin\c{c}li, S., Henkel, N., Ates, C., et al.
{\em Phys. Rev. Lett.} {\bf 107}, 153001 (2011).

\bibitem{low2012}
L\"ow, R., Weimer, H., Nipper, J., et al.
{\em J. Phys. B: At. Mol. Opt. Phys.} {\bf 45}, 113001 (2012).

\bibitem{Ketterle1999}
Ketterle, W., Durfee, D., and Stamper-Kurn, D.
Bose-Einstein Condensation in Atomic Gases: Proceedings of the International
  School of Physics `Enrico Fermi' Course CXI chapter Making, probing and
  understanding Bose-Einstein condensates, pp. 67 -- 176
IOS Press (1999).

\bibitem{Ketterle2008}
Ketterle, W. and Zwierlein, M.~W.
Ultra-Cold Fermi Gases: Proceedings of the International School of Physics
  "Enrico Fermi", Course ClXIV chapter Making, probing and understanding
  ultracold Fermi gases, pp. 95 -- 287
IOS Press, Amsterdam (2008).

\bibitem{Sassmannshausen2013}
Sa\ss{}mannshausen, H., Merkt, F., and Deiglmayr, J.
{\em Phys. Rev. A} {\bf 87}, 032519 (2013).

\bibitem{OSullivan1985}
O'Sullivan, M.~S. and Stoicheff, B.~P.
{\em Phys. Rev. A} {\bf 31}, 2718--2720 (1985).

\bibitem{Beterov2009}
Beterov, I.~I., Ryabtsev, I.~I., Tretyakov, D.~B., et al.
{\em Phys. Rev. A} {\bf 79}, 052504 (2009).

\bibitem{vogt2006}
Vogt, T., Viteau, M., Zhao, J., et al.
{\em Phys. Rev. Lett.} {\bf 97}, 083003 (2006).

\bibitem{westermann2006}
Westermann, S., Amthor, T., {de Oliveira}, A.~L., et al.
{\em The European Physical Journal D - Atomic, Molecular, Optical and Plasma
  Physics} {\bf 40}, 37--43 (2006)
10.1140/epjd/e2006-00130-3.

\bibitem{ryabtsev2010}
Ryabtsev, I.~I., Tretyakov, D.~B., Beterov, I.~I., et al.
{\em Phys. Rev. Lett.} {\bf 104}, 073003 (2010).

\bibitem{nipper2012}
Nipper, J., Balewski, J.~B., Krupp, A.~T., et al.
{\em Phys. Rev. X} {\bf 2}, 031011 (2012).

\bibitem{gurian2012}
Gurian, J.~H., Cheinet, P., Huillery, P., et al.
{\em Phys. Rev. Lett.} {\bf 108}, 023005 (2012).

\bibitem{mulken2007}
M{\"u}lken, O., Blumen, A., Amthor, T., et al.
{\em Phys. Rev. Lett.} {\bf 99}, 090601 (2007).

\bibitem{dieckmann1998}
Dieckmann, K., Spreeuw, R. J.~C., Weidem\"uller, M., et al.
{\em Phys. Rev. A} {\bf 58}, 3891 (1998).

\bibitem{schoser2002}
Schoser, J., Bat\"ar, A., L\"ow, R., et al.
{\em Phys. Rev. A} {\bf 66}, 023410 (2002).

\bibitem{catani2006}
Catani, J., Maioli, P., Sarlo, L.~D., et al.
{\em Phys. Rev. A} {\bf 73}, 033415 (2006).

\bibitem{chaudhuri2006}
Chaudhuri, S., Roy, S., and Unnikrishnan, C.~S.
{\em Phys. Rev. A} {\bf 74}, 023406 (2006).

\bibitem{dubessy2012}
Dubessy, R., Merloti, K., Longchambon, L., et al.
{\em Phys. Rev. A} {\bf 85}, 013643 (2012).

\bibitem{altin2010}
Altin, P.~A., Robins, N.~P., D\"oring, D., et al.
{\em Rev. Sci. Inst.} {\bf 81}, 063103 (2010).

\bibitem{lin2009}
Lin, Y.-J., Perry, A.~R., Compton, R.~L., et al.
{\em Phys. Rev. A} {\bf 2009}, 063631 (2009).

\bibitem{Fortagh2007}
Fort\'agh, J. and Zimmermann, C.
{\em Rev. Mod. Phys.} {\bf 79}, 235--289 (2007).

\bibitem{Reichel2011}
Reichel, J. and Vuletic, V. (2011)
Atom Chips,
Wiley-VCH Verlag GmbH \& Co. KGaA, .

\bibitem{tobepublished}
{Hofmann, C. S. et al.}
{\em To be published} (2013).

\bibitem{tiecke2009b}
Tiecke, T.~G., Gensemer, S.~D., Ludewig, A., et al.
{\em Phys. Rev. A} {\bf 80}, 013409 (2009).

\bibitem{Note1}
We use ALVASOURCES from Alvatec, which are chromate-free metal vapor sources of
  the type AS-3-Rb87(98\%)-20-F.

\bibitem{gotz2012}
G\"otz, S., H\"oltkemeier, B., Hofmann, C.~S., et al.
{\em Rev. Sci. Inst.} {\bf 83}, 073112 (2012).

\bibitem{jacob2011}
Jacob, D., Mimoun, E., Sarlo, L.~D., et al.
{\em New Journal of Physics} {\bf 13}, 065022 (2011).

\bibitem{clement2009}
Cl{\'e}ment, J.~F., Brantut, J.~P., {Robert-de-Saint-Vincent}, M., et al.
{\em Phys. Rev. A} {\bf 79}, 061406 (2009).

\bibitem{weber2002}
Weber, T., Herbig, J., Mark, M., et al.
{\em Science} {\bf 299}, 232 (2002).

\bibitem{zaiser2011}
Zaiser, M., Hartwig, J., Schlippert, D., et al.
{\em Phys. Rev. A} {\bf 83}, 035601 (2011).

\bibitem{kuppens2000}
Kuppens, S. J.~M., Corwin, K.~L., Miller, K.~W., et al.
{\em Phys. Rev. A} {\bf 62}, 013406 (2000).

\bibitem{townsend1996}
Townsend, C.~G., Edwards, N.~H., K.P.Zetie, et al.
{\em Phys. Rev. A} {\bf 53}, 1702 (1996).

\bibitem{ohara2001}
{O'Hara}, K.~M., Gehm, M.~E., Granade, S.~R., et al.
{\em Phys. Rev. A} {\bf 64}, 051403 (2001).

\bibitem{Lauber2011}
Lauber, T., K\"uber, J., Wille, O., et al.
{\em Phys. Rev. A} {\bf 84}, 043641 (2011).

\bibitem{Tauschinsky2010}
Tauschinsky, A., Thijssen, R. M.~T., Whitlock, S., et al.
{\em Phys. Rev. A} {\bf 81(6)}, 063411 (2010).

\bibitem{Hattermann2012}
Hattermann, H., Mack, M., Karlewski, F., et al.
{\em Phys. Rev. A} {\bf 86}, 022511 (2012).

\bibitem{Fleischhauer2005}
Fleischhauer, M., Imamoglu, A., and Marangos, J.
{\em Reviews of Modern Physics} {\bf 77(2)}, 633 (2005).

\bibitem{Pritchard2010}
Pritchard, J.~D., Maxwell, D., Gauguet, A., et al.
{\em Phys. Rev. Lett.} {\bf 105}, 193603 (2010).

\bibitem{Schempp2010}
Schempp, H., G\"unter, G., Hofmann, C.~S., et al.
{\em Phys. Rev. Lett.} {\bf 104(17)}, 173602 (2010).

\bibitem{sevincli2011}
Sevin\c{c}li, S., Ates, C., Pohl, T., et al.
{\em J. Phys. B} {\bf 44}, 184018 (2011).

\bibitem{Dicke1954}
Dicke, R.~H.
{\em Phys. Rev.} {\bf 93}, 99--110 (1954).

\bibitem{Heidemann2007}
Heidemann, R., Raitzsch, U., Bendkowsky, V., et al.
{\em Phys. Rev. Lett.} {\bf 99}, 163601 (2007).

\bibitem{Gaetan2009}
Ga{\"e}tan, A., Miroshnychenko, Y., Wilk, T., et al.
{\em Nature Physics} {\bf 5}, 115 (2009).

\bibitem{Urban2009}
Urban, E., Johnson, T.~A., Henage, T., et al.
{\em Nature Physics} {\bf 5}, 110 (2009).

\bibitem{Dudin2012b}
Dudin, Y.~O., Li, L., Bariani, F., et al.
{\em Nature Physics} {\bf 8}, 790 (2012).

\bibitem{Tong2004}
Tong, D., Farooqi, S.~M., Stanojevic, J., et al.
{\em Phys. Rev. Lett.} {\bf 93(6)}, 063001 (2004).

\bibitem{Singer2004}
Singer, K., Reetz-Lamour, M., Amthor, T., et al.
{\em Phys. Rev. Lett.} {\bf 93(16)}, 163001 (2004).

\bibitem{Vogt2007}
Vogt, T., Viteau, M., Chotia, A., et al.
{\em Phys. Rev. Lett.} {\bf 99}, 073002 (2007).

\bibitem{Reinhard2008}
Reinhard, A., Younge, K.~C., and Raithel, G.
{\em Phys. Rev. A} {\bf 78}, 060702(R) (2008).

\bibitem{Viteau2012}
Viteau, M., Huillery, P., Bason, M.~G., et al.
{\em Phys. Rev. Lett.} {\bf 109}, 053002 (2012).

\bibitem{Robinson2000}
Robinson, M.~P., Tolra, B.~L., Noel, M.~W., et al.
{\em Phys. Rev. Lett.} {\bf 85}, 4466 (2000).

\bibitem{Killian2007}
Killian, T.~C.
{\em Science} {\bf 316}, 705 (2007).

\bibitem{Li2004}
Li, W., Noel, M.~W., Robinson, M.~P., et al.
{\em Phys. Rev. A} {\bf 70}, 042713 (2004).

\bibitem{walz-flannigan2004}
Walz-Flannigan, A., Guest, J.~R., Choi, J.-H., et al.
{\em Phys. Rev. A} {\bf 69}, 063405 (2004).

\bibitem{Morrison2008}
Morrison, J.~P., Rennick, C.~J., Keller, J.~S., et al.
{\em Phys. Rev. Lett.} {\bf 101}, 205005 (2008).

\bibitem{Remington1999}
Remington, B.~A., Arnett, D., Paul, R., et al.
{\em Science} {\bf 284}, 1488--1493 (1999).

\bibitem{Horn1991}
Horn, H. M.~V.
{\em Science} {\bf 252}, 384 (1991).

\bibitem{Shuryak2009}
Shuryak, E.
{\em Progress in Particle and Nuclear Physics} {\bf 62}, 48 -- 101 (2009).

\bibitem{Ichimaru1982}
Ichimaru, S.
{\em Rev. Mod. Phys.} {\bf 54}, 1017--1059 (1982).

\bibitem{Killian2007b}
Killian, T.~C., Pattard, T., Pohl, T., et al.
{\em Physics Reports} {\bf 449}, 77 (2007).

\bibitem{Simien2004}
Simien, C.~E., Chen, Y.~C., Gupta, P., et al.
{\em Phys. Rev. Lett.} {\bf 92}, 143001 (2004).

\bibitem{Cummings2005}
Cummings, E.~A., Daily, J.~E., Durfee, D.~S., et al.
{\em Phys. Rev. Lett.} {\bf 95}, 235001 (2005).

\bibitem{Kuzmin2002}
Kuzmin, S.~G. and O'Neil, T.~M.
{\em Phys. Rev. Lett.} {\bf 88}, 065003 (2002).

\bibitem{Pohl2004}
Pohl, T., Pattard, T., and Rost, J.~M.
{\em Phys. Rev. A} {\bf 70}, 033416 (2004).

\bibitem{Bergeson2011}
Bergeson, S.~D., Denning, A., Lyon, M., et al.
{\em Phys. Rev. A} {\bf 83}, 023409 (2011).

\bibitem{Beterov2007}
Beterov, I.~I., Tretyakov, D.~B., Ryabtsev, I.~I., et al.
{\em Phys. Rev. A} {\bf 75}, 052720 (2007).

\bibitem{Barbier1987}
Barbier, L. and Cheret, M.
{\em Journal of Physics B Atomic Molecular Physics} {\bf 20}, 1229--1248
  (1987).

\bibitem{Kumar1999}
Kumar, A., Sahaa, B.~C., Weatherforda, C.~A., et al.
{\em Journal of Molecular Structure Theochem} {\bf 487}, 1 (1999).

\bibitem{Murillo2001}
Murillo, M.~S.
{\em Phys. Rev. Lett.} {\bf 87}, 115003 (2001).

\bibitem{Shukla2011}
Shukla, P.~K. and Avinash, K.
{\em Phys. Rev. Lett.} {\bf 107}, 135002 (2011).

\bibitem{Mandel1979}
Mandel, L.
{\em Optics Letters} {\bf 4}, 205--207 (1979).

\bibitem{Wuster2010}
W\"uster, S., Stanojevic, J., Ates, C., et al.
{\em Phys. Rev. A} {\bf 81}, 023406 (2010).

\bibitem{Breyel2012}
Breyel, D., Schmidt, T.~L., and Komnik, A.
{\em Phys. Rev. A} {\bf 86}, 023405 (2012).

\bibitem{Garttner2012b}
G\"arttner, M., Heeg, K.~P., Gasenzer, T., et al.
{\em arXiv:1203.2884} (2012).

\bibitem{Fleischhauer2000}
Fleisch\-hauer, M. and Lukin, M.~D.
{\em Phys. Rev. Lett.} {\bf 84}, 5094 (2000).

\bibitem{Ates2006}
Ates, C., Pohl, T., Pattard, T., et al.
{\em J. Phys. B} {\bf 39}, L233 (2006).

\bibitem{Petrosyan2011}
Petrosyan, D., Otterbach, J., and Fleischhauer, M.
{\em Phys. Rev. Lett.} {\bf 107}, 213601 (2011).

\bibitem{Parigi2012}
Parigi, V., Bimbard, E., Stanojevic, J., et al.
{\em Phys. Rev. Lett.} {\bf 109}, 233602 (2012).

\bibitem{Shahmoon2011}
Shahmoon, E., Kurizki, G., Fleischhauer, M., et al.
{\em Phys. Rev. A} {\bf 83}, 033806 (2011).

\bibitem{Gorshkov2011}
Gorshkov, A.~V., Otterbach, J., Fleischhauer, M., et al.
{\em Phys. Rev. Lett.} {\bf 107}, 133602 (2011).

\bibitem{Fleischhauer2008}
Fleischhauer, M., Otterbach, J., and Unanyan, R.~G.
{\em Phys. Rev. Lett.} {\bf 101}, 163601 (2008).

\bibitem{Nikoghosyan2012}
Nikoghosyan, G., Zimmer, F.~E., and Plenio, M.~B.
{\em Phys. Rev. A} {\bf 86}, 023854 (2012).

\bibitem{Honer2011}
Honer, J., L\"ow, R., Weimer, H., et al.
{\em Phys. Rev. Lett.} {\bf 107}, 093601 (2011).

\bibitem{Stanojevic2012}
Stanojevic, J., Parigi, V., Bimbard, E., et al.
{\em Phys. Rev. A} {\bf 86}, 021403 (2012).

\bibitem{Friedler2005}
Friedler, I., Petrosyan, D., Fleischhauer, M., et al.
{\em Phys. Rev. A} {\bf 72}, 043803 (2005).

\bibitem{Petrosyan2008}
Petrosyan, D. and Fleischhauer, M.
{\em Phys. Rev. Lett.} {\bf 100}, 170501 (2008).

\bibitem{Gunter2012}
G\"unter, G., {Robert-de-Saint-Vincent}, M., Schempp, H., et al.
{\em Phys. Rev. Lett.} {\bf 108}, 013002 (2012).

\end{thebibliography}

\end{document}